\documentclass[aps,superscriptaddress,amsmath,twocolumn,amssymb,floatfix,showpacs,superscriptaddress,nofootinbib,longbibliography]{revtex4-1}
\usepackage{MnSymbol}
\usepackage{braket}
\usepackage[dvipsnames]{xcolor}
\usepackage{float}
\usepackage{subfigure}
\usepackage{tikz}
\usepackage[colorlinks=true,linktoc=page,linkcolor=OliveGreen,citecolor=purple,urlcolor=violet]{hyperref}

\mathchardef\mhyphen="2D 

\newcommand{\ie}{{i.e.,\,\,}}
\newcommand{\eg}{{e.g.,~}}

\newcommand\bea{\begin{eqnarray}}
\newcommand\eea{\end{eqnarray}}
\newcommand\beq{\begin{equation}}  
\newcommand\eeq{\end{equation}}

\usepackage[normalem]{ulem}
\definecolor{lime}{HTML}{A6CE39}
\usepackage{sidecap,tikz}
\DeclareRobustCommand{\orcidicon}{\hspace{-1.0mm}
	\begin{tikzpicture}
		\draw[lime, fill=lime] (0.0,0.0) 
		circle [radius=0.15] 
		node[white] {{\fontfamily{qag}\selectfont \tiny \,ID}};
		\draw[white, fill=white] (-0.0525,0.095) 
		circle [radius=0.007];
	\end{tikzpicture}
	\hspace{-3.0mm}
}
\foreach \x in {A, ..., Z}{\expandafter\xdef\csname orcid\x\endcsname{\noexpand\href{https://orcid.org/\csname orcidauthor\x\endcsname}{\noexpand\orcidicon}}
}

\AtBeginDocument{%
	\newwrite\bibnotes
	\def\bibnotesext{Notes.bib}
	\immediate\openout\bibnotes=\jobname\bibnotesext
	\immediate\write\bibnotes{@CONTROL{REVTEX41Control}}
	\immediate\write\bibnotes{@CONTROL{%
			apsrev41Control,author="08",editor="1",pages="1",title="1",year="1"}}
	\if@filesw
	\immediate\write\@auxout{\string\citation{apsrev41Control}}%
	\fi
}%

\begin{document}

\title{Tailoring topological band properties of twisted double bilayer graphene: effects due to spin-orbit coupling}  

\author{Kamalesh Bera\orcidA{}}
\email{kamalesh.bera@iopb.res.in}
\affiliation{Institute of Physics, Sachivalaya Marg, Bhubaneswar-751005, India}
\affiliation{Homi Bhabha National Institute, Training School Complex, Anushakti Nagar, Mumbai 400094, India}

\author{Priyanka Mohan\orcidB{}}
\email{priya11198@gmail.com}
\affiliation{Dolat Capital Market Pvt Ltd, India}

\author{Arijit Saha\orcidC{}}
\email{arijit@iopb.res.in}
\affiliation{Institute of Physics, Sachivalaya Marg, Bhubaneswar-751005, India}
\affiliation{Homi Bhabha National Institute, Training School Complex, Anushakti Nagar, Mumbai 400094, India}

\begin{abstract}
Our theoretical study unfolds the topological phase transitions (within bands of the Moir\'e super-lattice) in small angle twisted double bilayer graphene (tDBLG) under the influence of external gate voltage and intrinsic spin-orbit coupling (SOC) for both AB-AB and AB-BA stacking configurations. Utilizing a low-energy continuum model, we investigate the band structure and perform a comprehensive topological characterization of the system by analysing the direct band gap closing as well as various Chern numbers. In the absence of SOC, the tDBLG exhibits characteristics of a valley Hall insulator. However, in the presence of SOC, we observe a transition to a quantum spin Hall insulator state and band topology emerges in the parameter spaces of non-topological regime without SOC. Furthermore, we conduct a comparative analysis between untwisted double bilayer graphene and tDBLG to assess the impact of twisting on the system's properties. Our findings reveal the construction of topological phase diagrams that showcase distinct phases arising from changes in the twist angle compared to the untwisted case. These phase diagrams provide valuable insights into the diverse topological phases achievable in tDBLG with SOC. Our findings contribute to the understanding of the interplay between small twist angle, SOC, and external electric field on the topological band properties of twisted multilayer graphene systems.
\end{abstract}

\maketitle

\section{Introduction}

Graphene is a two-dimensional (2D) sheet of carbon atoms, which has been investigated extensively in the past few decades~\cite{Geim2007, RMP-Graphene-review}. Also, stacking of such 2D sheets, the so-called van-der Waals materials (e.g. bilayer~\cite{McCann_2013-BilayerReview}, trilayer, multilayer~\cite{Multilayer} etc.), have been a research field of interest for a long time. However, the introduction of twist (rotational misalignment) between such 2D-sheets of van-der Waals materials~\cite{Santos-Peres-tBLG, Shallcross-tBLG} and their experimental realisation~\cite{Li2010-vanHoveSingularity} 
has opened up a new frontier of research in the field of quantum condensed matter physics.   

Interestingly, in twisted bilayer graphene (tBLG), a long period  Moir\'e pattern emerges in the real space geometry and in reciprocal space the Dirac cones get modified and as a result, the highly dispersive low energy bands (Dirac cones near the two valleys) of monolayer graphene become flat~\cite{MacDonald-tBLG, Moon-tBLG, Koshino-tBLG}. In recent times, the experimental observation of the unconventional superconductivity and correlated insulating phase~\cite{Cao2018-corr_insulator, Cao2018-unconv_sc} at the magic angle tBLG has triggered a huge amount of attention from both 
theoretical~\cite{MI+SC-tBLG, All_Magic_Angle-tBLG, Origin_of_Magic_Angle-tBLG, CIS-tBLG, Magnetism-tBLG, Heavy_fermion-tBLG,tbg-Constantinos} and experimental~\cite{Wu2021-chernIns-expt, 
Nuckolls2020-tBLG(xpt), Oh2021-tBLG(xpt), SOC-tBLG1,SOC-tBLG2} perspectives. Nevertheless, not only the graphene bilayer, but depending on the number of layers, stacking geometry, the materials used, and their lattice structure one can have a plethora of class of the so-called Moir\'e materials. For \eg twisted monolayer, bilayer graphene~\cite{Mono-Bi-graphene1, Mono-Bi-graphene2}, twisted trilayer graphene~\cite{Park2021-tTLG1, Kim2022-tTLG2}, twisted double bilayer graphene (tDBLGs)~\cite{Liu2020-tDBLG1,tDBLG2, He2021-tDBLG3, Cao2020-tDBLG4, Haddadi2020-tDBLG5, Adak2022-tDBLG6, Zhang2021-tDBLG7,Chakraborty_2022,Sinha2022}, transition-metal dichalcogenide (TMD) homobilayers~\cite{TMD-homobilayers1, TMD-homobilayers2},  rhombohedral-stacked trilayer graphene (rTLG) aligned with hexagonal boron nitride (hBN)~\cite{rTTLG+hBN1,rTTLG+hBN2} etc. have gained significant research endeavour.

Here, in this work, we consider tDBLG which is made up of stacking of two bi-layer graphenes with a relative twist angle between them~\cite{Chebrolu, Koshino-tDBLG}. We also assume that each layer  possesses an intrinsic spin-orbit coupling (SOC). As we know, the presence of spin-orbit interaction in monolayer graphene turns it from a semimetal to a quantum spin-Hall insulator (QSHI)~\cite{Kane-mele_QSH1, Kane-mele_QSH2} \ie a 2D topological insulator (TI). Thus one intriguing aspect remains to explore is the interplay between the twist and the spin-orbit interaction~\cite{SOC-tBLG1, SOC-tBLG2}. However there exists very few works in this direction especially from the topological band point of view in tDBLG, thus it motivates to explore this aspect in our work. Also, it is well known that 
graphene itself exhibits vanishingly small intrinsic SOC. However, there are monolayers of silicene, germanene, stranene~\cite{Ezawa_2012-silicene} etc. with hexagonal lattice structure that contain significant intrinsic SOC and our model can mimic twisted double bilayer systems which are made 
up of such monolayers.

In literature, the electronic band structure and topological band properties of tDBLG have been studied previously~\cite{SOC-tBLG1, SOC-tBLG2}. One key difference between tBLG and tDBLG is that while 
the application of a transverse electric field doesn't open up a band gap near the charge neutrality point in tBLG, it does open up a band gap in case of tDBLG~\cite{tBLG-Gate-voltage, Koshino-tDBLG}. 
Also, it has been found that tDBLG undergoes different topological phase transitions [quantum valley-Hall insulator (QVHI)] as the gate-voltage is tuned~\cite{Koshino-tDBLG, Chebrolu, Mohan2021}. Thus considering SOC in our model adds an extra knob to control the topological electronic band properties of such systems.

In this article, we emphasize the effect of intrinsic SOC on the electronic band properties of tDBLG. 
In the presence of SOC, we observe that the first valence band and conduction band of the low energy tDBLG band structure become gapped, while they appear gapless in the absence of intrinsic SOC. 
On the other hand, for some other parameter values, as a result of the interplay between the SOC and gate voltage, it has been observed that the above-mentioned bands become gapless, from being gapped without SOC. In the latter part of our work, we attempt to understand the topological properties of the tBLG band structure. First, we look at the direct band-gap closing as a function of gate voltage in the presence of a fixed intrinsic-SOC strength. We observe that a new type of gap-closing takes place due to the presence of the SOC. This is absent when SOC is not incorporated. 
The band-gap in the Moir\'e Brillouin zone (mBZ) exhibits Lifshitz-like transitions~\cite{McCann_2013-BilayerReview} (\ie in bilayer graphene, the presence of trigonal warping term, breaks the circular 
iso-energetic line around a Dirac point into four different pockets of triangular shape, one around the central Dirac point and the other three around satellite Dirac points). We further show that indeed the gap closings correspond to topological phase transitions. However, at the extra direct band-gap closing point, we observe that the band gap which closes at each satellite Dirac point, further takes a rod-like shape and the new gap closing points are formed on the other side of the dimers. Furthermore, we compute the total Chern number upto the valence band and investigate the effect of enhanced SOC strength on the phase diagrams.
Moreover, we show the topological phase diagrams for all four different sectors of the valley and spin 
considering both AB-AB and AB-BA stacked tDBLG and calculate the valley Chern number, spin Chern number, spin-valley Chern number, and the total Chern number to emphasize the appropriate topological characterization of the system. We find that while in the absence of intrinsic SOC, 
the tDBLG becomes a QVHI in certain parameter regime of gate voltage and twist angle, they also become a quantum spin-Hall insulator (QSHI) in the presence of SOC in certain other parameter regime. 
Finally, we provide a comparison between the topological band properties of untwisted double-bilayer graphene (uDBLG) and tDBLG to highlight the role of twist angle in presence of SOC.

The remainder of the article is organised as follows. In Sec.~\ref{Sec:II}, we introduce the model Hamiltonian for uDBLG and show how the model Hamiltonian for tDBLG is constructed in the presence 
of both external gate-voltage (electric field) and intrinsic-SOC. 
In Sec.~\ref{Sec:III}, we show the electronic band structure of tDBLG for different values of the gate voltages and intrinsic-SOC strengths. Sec.~\ref{Sec:IV} is devoted to the discussion of direct 
band-gap closings as a function of the electric field for a particular twist angle and SOC strength. 
Afterwards, in Sec.~\ref{Sec:V}, we discuss various topological phase diagrams to emphasize the 
effect of enhanced SOC strength on the band properties and comparison with the untwisted case.
In Sec.~\ref{Sec:VI}, we discuss unusual band-gap closings in mBZ in the form of Lifshitz-like 
transition. Finally, we summarize and conclude our paper in Sec.~\ref{Sec:VII}.

\section{Model and Method}\label{Sec:II}
In this section, we discuss how to construct the model Hamiltonian for tDBLG in presence of small
twist angle, transverse electric field and SOC. 
\subsection{Model Hamiltonian}\label{subsec:IIA}
When two layers of graphene are stacked on top of each other and slightly rotated, it forms a tBLG. 
Now, in place of the single-layer graphene, if one considers stacking of slightly rotated bilayer graphenes, then this forms a tDBLG. Below we describe the low energy continuum model for the tDBLG. Before that, we write the low energy continuum model for bilayer graphene and discuss the construction of the tDBLG Hamiltonian therein. The low energy continuum Hamiltonian for a AB-stacked 
and BA-stacked bilayer graphene near valley-$K$ in the basis of $(A_{1},B_{1},A_{2},B_{2})$ can be written as,

\begin{align}
H^{\rm{AB}}_{\bf k}= 
\left( \begin{array}{cc}
h_{\bf k} & g^\dagger_{\bf k}\\
g_{\bf k} & h^{\prime}_{\bf k} \end{array}\right)\ ,~
H^{\rm{BA}}_{\bf k}= 
\left( \begin{array}{cc}
h^{\prime}_{\bf k} & g_{\bf k}\\
g^\dagger_{\bf k} & h_{\bf k} \end{array}\right)\ ,
\label{Eq:H:BLG1}
\end{align}	

where,
\begin{eqnarray} 
h_{ \bf k} &=
\left( \begin{array}{cc}
0  & -\hbar v k_- \\
-\hbar v k_+ & \Delta'
\end{array}\right)\ ,~
h'_{\bf k} 
&=
\left( \begin{array}{cc}
\Delta'  & -\hbar v k_- \\
-\hbar v k_+ & 0
\end{array}\right)\ , \nonumber\\
&g_{ \bf k} 
=&\!\!\!\!\!\!\!\!\!\!\!\!\!\!\!\!\!\!\!\!\!\!\!\!\!\!\!\!\!\!\!\!
\left( \begin{array}{cc}
\hbar v_4 k_+  & \gamma_1 \\
\hbar v_3 k_-  & \hbar v_4 k_+
\end{array}\right)\ .
\label{Eq:H:BLG:11}
\end{eqnarray}	

\begin{figure}[h]
	\centering
	\subfigure{\includegraphics[width=0.48\textwidth]{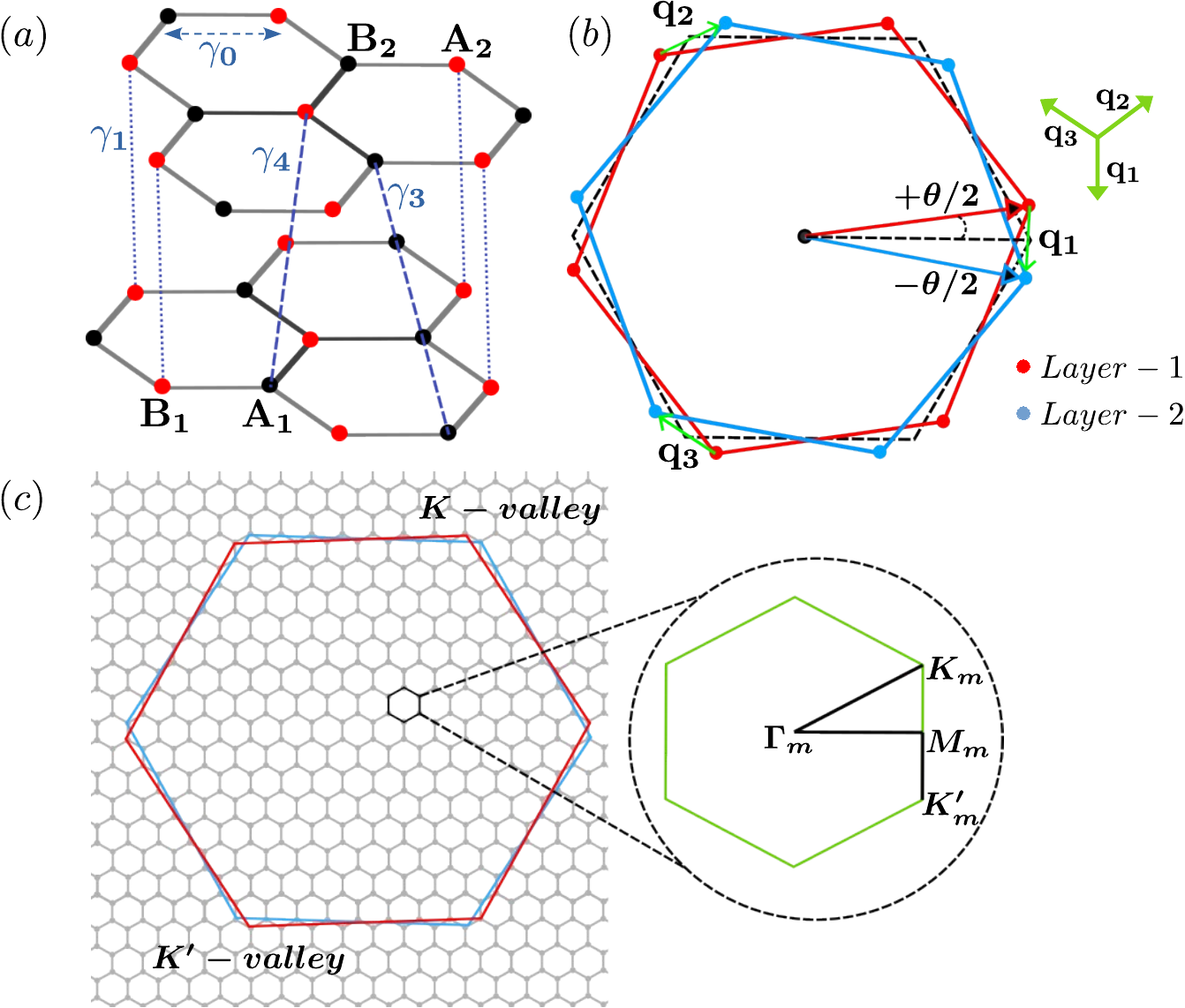}}
	\caption{(a) Schematic of bilayer graphene where, $(A_{1},B_{1}), (A_{2}, B_{2})$ denote the 
A, B sublattices of the layer-1 and layer-2 respectively. Here, $\gamma_{0}$, $\gamma_{1}$,
$\gamma_{3}$, $\gamma_{4}$ represent different hopping elements between sublattices of the same layer and different layers. (b) The red and cyan hexagons denote the Brillouin zone of the two bilayer graphenes rotated centering each other by $+\theta$/2 and $-\theta$/2. Here, $\mathbf{q_{1}}$, 
$\mathbf{q_{2}}$, $\mathbf{q_{3}}$ form the mBZ. (c) The smaller hexagonal tiling is the folded Brillouin zone \ie the mBZ. The larger two hexagons correspond to the Brillouin zone of the bilayers. 
One of the smaller hexagons (the mBZ) is zoomed in for clarity and shown with the high symmetry 
path along which the band structure is calculated.
}
	\label{mBZ2}
\end{figure}

Here, $k_{\pm} = \xi k_{x} \pm i k_{y} $ with $\xi$ takes value $\pm 1$ for valley-$K$ and valley-$K^{\prime}$ respectively. Also, $h_{\bf k}$ and $h'_{\bf k}$ denote the low energy model Hamiltonians for each single layer of graphene, with $\Delta'$ being the onsite energy of dimer sites. Here, $\Delta'$ accounts for the distinction between dimer (sites in the bilayer that are exactly on top of each other) and non-dimer (sites in the bilayer that don't have any site exactly on top of it) and $v$ is the group velocity of electrons in graphene band. $g_{\bf k}$ represents the interlayer coupling between the two single 
layers. Furthermore, $\gamma_1$ in $g_{\bf k}$ [see Eq.~(\ref{Eq:H:BLG:11})] provides the coupling between the dimer sites, whereas $\gamma_3$, $\gamma_4$ represent coupling within non-dimer sites.
These couplings are schematically shown in Fig.~\ref{mBZ2}(a). Here, in our calculations, we  consider, $\hbar v/a =2135.4$ meV~\cite{RMP-Graphene-review,McCann_2013-BilayerReview,Koshino-tBLG} (with $a = 0.246$ nm being the lattice constant of graphene),  $\Delta' = 50$ meV, $\gamma_1 = 400$ meV, $\gamma_3 = 320 $ meV, $\gamma_4 = 44$ meV~\cite{McCann_2013-BilayerReview,Koshino-tBLG,McCann_2011-BilayerReview}. Also, $v_{i}$ is related to $\gamma_{i}$ 
as follows, $v_{i} = \sqrt{3}\gamma_{i} a/2\hbar$ (with $i = 3,4$). In bilayer graphene, $v_3$ and $v_4$ correspond to the trigonal warping and the particle-hole asymmetry term respectively.

Depending on the stacking of the bilayers, the double bilayer can exhibit two stable configurations: 
one is AB-AB and the other one is AB-BA. Here we write the low energy model Hamiltonian for uDBLGs~\cite{Multilayer} near valley-$K$ and in the basis $(A_{1},B_{1},A_{2},B_{2},A_{3},B_{3},A_{4},B_{4})$.

\begin{eqnarray}	
H^{\rm{AB\!-\!AB}}_{k} = 
\left( \begin{array}{cccc}
	h_{k} & g^\dagger_{k} & 0 & 0\\
	g_{k} & h'_{k} & g^\dagger_{k} & 0\\
	0 & g_{k} & h_{k} & g^\dagger_{k}\\
	0 & 0 & g_{k} & h'_{k}\\
	\end{array}\right)\ ,
\label{Eq:H:DBLG1}
\end{eqnarray}

\begin{eqnarray}
H^{\rm{AB\!-\!BA}}_{k}= 
\left( \begin{array}{cccc}
h_{k}  &  g^\dagger_{k}  &  0  &  0\\
g_{k}  &  h'_{k}  &  g^\dagger_{k}  &  0\\
0  &  g_{k}  &  h'_{k}  &  g_{k}\\
0  &  0  &  g^\dagger_{k}  &  h_{k}\\
\end{array}\right)\ ,
\label{Eq:H:DBLG2}
\end{eqnarray}

Afterwards, we discuss the construction of the low-energy continuum model Hamiltonian for tDBLG. The procedure is almost the same as the path followed to construct the low-energy continuum Hamiltonian for tBLG~\cite{MacDonald-tBLG,Moon-tBLG,Koshino-tBLG}. Here, we consider the first bilayer is rotated by $-\theta/2$ and the second bilayer follows a rotation of $\theta/2$ with respect 
to each other.  

Likewise tBLG, in tDBLG the effect of twist appears in similar ways~\cite{Koshino-tDBLG,Chebrolu}. There are mainly two agencies through which the effect of twist is considered in the low energy continuum model of tDBLG, (i) each low energy bilayer Hamiltonian becomes $\theta$ dependent, and (ii) the tunneling between the two bilayers gets modified. On account of the first case, one 
can simply write, 

\begin{eqnarray}
H^{AB}_{\bf k}(\theta)=\left(\begin{array}{cccc}%
		 0  & -\hbar v k_{-}^{\theta} & \hbar v_4 k_{-}^{\theta}& \hbar v_3 k_{+}^{\theta}\\
		 -\hbar v k_{+}^{\theta} & \Delta' & \gamma_1 &\hbar v_4 k_{-}^{\theta} \\
		 \hbar v_4 k_{+}^{\theta}  & \gamma_1 &\Delta'&-\hbar v k_{-}^{\theta}\\
		 \hbar v_3 k_{-}^{\theta}  & \hbar v_4 k_{+}^{\theta} & -\hbar v k_{+}^{\theta} & 0\\
	\end{array}\right)\ ,
	\label{Eq:H:AB-AB-Full}
\end{eqnarray}
where, $\mathbf{k^{\theta}} = R(\theta) \mathbf{k}$ with $R(\theta) $ being the rotation matrix in 2D.

Then, the other part is the interlayer tunneling and the derivation for which has already been presented in the literature~\cite{MacDonald-tBLG, Moon-tBLG}. Here, we write down that expression and 
 try to understand the terms from the Brillouin zone (BZ) picture. In Fig.~\ref{mBZ2}(b), we show two Hexagonal BZs corresponding to layer-1 (red color) and layer-2 (cyan color) of bilayers, 
rotated by $\theta/2$ and $-\theta/2$ with respect to each other. This misalignment of the BZs gives rise to three distinct vectors $\mathbf{q_{1}}$, $\mathbf{q_{2}}$, $\mathbf{q_{3}}$ (as shown in 
Fig.~\ref{mBZ2}(b) with the green arrows) which connect $K$/$K^{\prime}$ valley of BZs in the first and second layer. 
Here, the magnitude $\lvert q_{i=1,2,3}\rvert = (8\pi \sin{\theta/2})/3\sqrt{3}d$ (with $d$ being the lattice constant in a single layer of graphene) and directions of $\mathbf{q}$'s are given by, $\mathbf{q_{1}} \equiv (0,-1)$, $\mathbf{q_{2}} \equiv (\sqrt{3}/2,1/2)$, $\mathbf{q_{3}} \equiv (-\sqrt{3}/2,1/2)$. 

The effective interlayer coupling between the two layers (\ie tBLG) is given by the matrix $T$ as follows~\cite{MacDonald-tBLG, Moon-tBLG, Koshino-tDBLG},
\begin{eqnarray}
T = T_{1} + T_{2}\exp^{i\xi \mathbf{G_{1}^{m}} \cdot \mathbf{r}} + T_{3}\exp^{i\xi (\mathbf{G_{1}^{m}} + \mathbf{G_{2}^{m}}) \cdot \mathbf{r}}\ ,
\label{Eq:H:AB-AB-Full1}
\end{eqnarray}
where, $\mathbf{G^{m}} = n_{1}\mathbf{G_{1}^{m}} + n_{2}\mathbf{G_{2}^{m}}$ is the reciprocal lattice vector in the mBZ (the superscript $m$ in $\mathbf{G}$'s denotes that) and $n_{1}$, $n_{2}$ are integers. The tunneling matrices in Eq.~(\ref{Eq:H:AB-AB-Full1}) are given by

\begin{eqnarray}
T_{1} &=
\left( \begin{array}{cc}
u  & u^{\prime} \\
u^{\prime} & u
\end{array}\right) ~~,~~
T_{2} 
&=
\left( \begin{array}{cc}
u  & u^{\prime} \omega^{-\xi} \\
u^{\prime}\omega^{\xi} & u
\end{array}\right) ~~,~~ \nonumber\\
&T_{3} 
=&\!\!\!\!\!\!\!\!\!\!\!\!\!\!\!\!\!\!\!\!
\left( \begin{array}{cc}
u  & u^{\prime} \omega^{\xi} \\
u^{\prime}\omega^{-\xi} & u
\end{array}\right)\ .
\label{Eq:H:BLG:1}
\end{eqnarray}
Here, $u$ and $u^{\prime}$ stand for tunneling amplitudes between AA/BB and AB/BA sublattices respectively. We choose $u = 79.7$ meV, $u^{\prime} = 97.5$ meV~\cite{Moon-tBLG,Koshino-tBLG,corrugation-DFT-uu,Dai2016-uu} and $\omega = e^{2 \pi i/3}$ where $2 \pi/3$ denotes the 
angle between three base vectors of the mBZ.

Thus, the model Hamiltonian for the tDBLG can be written as,
\begin{widetext}
\begin{eqnarray}
H^{\rm{AB-AB}}_{\bf k}=\left(\begin{array}{cccc}%
	H^{AB}_{\bf k}(\theta/2) & \tilde{T_1}^\dagger & \tilde{T_2}^\dagger & \tilde{T_3}^\dagger\\
	\tilde{T_1} & H^{AB}_{\bf k+\bf q_1}(-\theta/2) & 0 & 0\\
	\tilde{T_2} & 0 & H^{AB}_{\bf k+\bf q_2}(-\theta/2)& 0\\
	\tilde{T_3} & 0 & 0 & H^{AB}_{\bf k+ \bf q_3}(-\theta/2) 
\end{array}\right)\ ,
\label{Eq:H:AB-AB-Full}
\end{eqnarray}
\end{widetext}
This Hamiltonian acts on the sixteen component spinor, 
$\Psi = (\psi_{0}, \psi_{1}, \psi_{2}, \psi_{3})$. Here, $\psi_{0}$ is a four component spinor 
near the Dirac point of layer-1 (\ie first bilayer) which is connected to $\psi_{i=1,2,3}$ (each of them four component spinor) 
near the Dirac point of layer-2 (\ie second bilayer) at momenta $q_{i}$'s via three distinct interlayer couplings $\tilde{T}_{i=1,2,3}$. These couplings read as
\begin{eqnarray}
\tilde{T_{i}} &=
\begin{pmatrix}
0  & 1 \\
0 & 0 \\
\end{pmatrix} \otimes T_i\ .
\label{Eq:H:BLG:2}
\end{eqnarray}

The above Hamiltonian is written within the four-wave approximations. However, to obtain the low energy bands, one needs to go beyond this approximation. As a result, a momentum space lattice
[as shown by the small hexagons in the background of Fig.~\ref{mBZ2}(c)] is generated with $\mathbf{q_{1}}, \mathbf{q_{2}},\mathbf{q_{3}}$ being the basis vectors. For that reason, one can also think that the above Hamiltonian [Eq.~(\ref{Eq:H:AB-AB-Full})] is written in the first mBZ. In our analysis, we consider upto the seventh nearest neighbor and construct the Hamiltonian 
in the corresponding basis to diagonalize it numerically. 

\subsection{Effect of external gate-voltage and presence of intrinsic SOC}\label{subsec:IIB}
Here, we incorporate the effect of the transverse electric field and the intrinsic SOC in tDBLG
Hamiltonian in the following way described below, 

First, we write the effect of a transverse electric field on the basis 
$(A_{1},B_{1},A_{2},B_{2},A_{3},B_{3},A_{4},B_{4})$,

\begin{eqnarray}
V = 
\left( \begin{array}{cccc}
\frac{3}{2} \Delta \mathbf{I_{2\times2}} &  0  &  0  &  0\\
0  &  \frac{1}{2} \Delta \mathbf{I_{2\times2}}  &  0  &  0\\
0  &  0  &  -\frac{1}{2} \Delta \mathbf{I_{2\times2}}  &  0\\
0  &  0  &  0  & -\frac{3}{2} \Delta \mathbf{I_{2\times2}}\\
\end{array}\right)\ ,
\label{Eq:H:Ez} 
\end{eqnarray}
where, $\Delta$ denotes the uniform potential difference between two consecutive layers. 

Consideration of spin degree of freedom doubles the size of the basis, \ie now the basis we have $(A_{1 \uparrow}, B_{1 \uparrow}, A_{2 \uparrow}, B_{2 \uparrow}, A_{3 \uparrow}, B_{3 \uparrow}, A_{4 \uparrow}, B_{4 \uparrow}, A_{1 \downarrow}, B_{1 \downarrow}, A_{2 \downarrow}, B_{2 \downarrow}, \newline A_{3 \downarrow}, B_{3 \downarrow}, A_{4 \downarrow}, B_{4 \downarrow})$. As we don't consider any spin flip scattering in our problem, the $z$-component of the spin $s_{z}$ is still a good quantum number 
and one can decompose the total Hamiltonian into block diagonal form for each spin-sector. Below we write the model Hamiltonian with intrinsic SOC for a particular spin sector as,

\begin{eqnarray}
L_{so}^{s_{z}} = 
\left( \begin{array}{cccc}
s_{z} \lambda_{so} \sigma_{z}  &  0  &  0  &  0\\
0  &  s_{z} \lambda_{so} \sigma_{z}   &  0  &  0\\
0  &  0  &  s_{z} \lambda_{so} \sigma_{z}  &  0\\
0  &  0  &  0  & s_{z} \lambda_{so} \sigma_{z} \\
\end{array}\right)\ ,
\label{Eq:H:Lso}
\end{eqnarray}
where, $\lambda_{so}$ is the intrinsic SOC strength, $\sigma_{z}$ is the Pauli matrix and $s_{z}$ = $\pm 1$ denotes the spin-up and spin-down sectors respectively. In principle, one
can also consider both the spin-flipping terms (such as Rashba SOC) and sub-lattice dependent intrinsic SOC in the model. However, within the context of our study (\ie for the understanding of new topological phases in twisted systems), we expect that Rashba SOC does not have a detrimental effect; it may change the boundaries of the topological phase diagrams. Nevertheless, it is not expected to introduce new phases. Thus, we neglect it in our model.

Finally, writing the transverse electric field (in terms of gate voltage) ($V$) and SOC ($L_{so}^{s_{z}}$) Hamiltonians in the same basis as Eq.~(\ref{Eq:H:AB-AB-Full}) and adding theses three matrices yield the following block Hamiltonians $H^{\rm{AB-AB}}(K;\uparrow)$ and $H^{\rm{AB-AB}}(K;\downarrow)$ for $s_{z}=\pm 1$ respectively. Similarly, one can have $H^{\rm{AB-AB}}(K^{\prime};\uparrow)$ and $H^{\rm{AB-AB}}(K^{\prime};\downarrow)$, when the tDBLG Hamiltonian [Eq.(\ref{Eq:H:AB-AB-Full})] is constructed around the $K^{\prime}$-valley. As there is no intervalley scattering present in our model Hamiltonian, valley contribution can also be written down separately. Following the changes in Eq.~(\ref{Eq:H:DBLG2}), one can construct the $\text{AB-BA}$ tDBLG Hamiltonian ($H^{\rm{AB-BA}}_{\bf k}$) in a similar fashion and finally the total Hamiltonian is composed of the four block Hamiltonians, $H^{\rm{AB-BA}}(K;\uparrow)$, $H^{\rm{AB-BA}}(K;\downarrow)$, $H^{\rm{AB-BA}}(K^{\prime};\uparrow)$,  $H^{\rm{AB-BA}}(K^{\prime};\downarrow)$ depending on the valley and spin degrees of freedoms. Also, similar decompositions hold for the untwisted case as well.

\section{Electronic band spectrum}\label{Sec:III}
In this section, we discuss the low-energy electronic band structure of both the untwisted and twisted double bilayer graphenes in the presence of intrinsic SOC.
\begin{figure}[h]
	\centering
	\subfigure{\includegraphics[width=0.4\textwidth,height=0.4\textwidth]{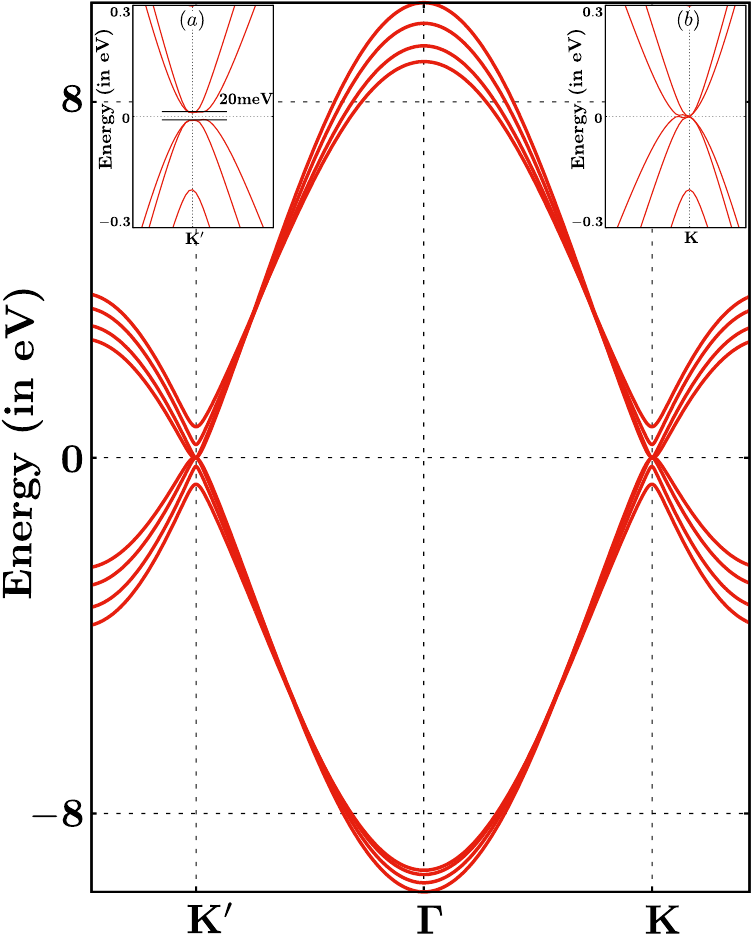}}
	\caption{The Band structure of uDBLG is depicted along the corner ($K$ and $K^{\prime}$) and center ($\Gamma$) of the hexagonal BZ. The inset (a) is highlighting the gapped
	spectrum near the valley-$K^{\prime}$, whereas the region near the valley-$K$ remains gapless as indicated by the inset (b) in the top right corner. This is shown for $s_{z} = +1$. 
	}
	\label{uDBLG(AB-AB)_bands}
\end{figure}
\subsection{Band dispersion of uDBLG}
Here in Fig.~\ref{uDBLG(AB-AB)_bands}, we show the band structure of the uDBLG for the spin-up sector (\ie $s_{z} = +1$). We use the same parameter ($v$, $\gamma_1$, $\gamma_3$, $\gamma_4$) values, mentioned in the previous section, for analysing the band structure. This band structure is depicted along the path $K^{\prime}$ - $\Gamma$ - $K$ in the original hexagonal BZ of graphene. In order to calculate the band structure, in Eq.~(\ref{Eq:H:DBLG1}) we replace $\hbar v k_{\pm}$ by $f_{p}$ and $f_{p}^{*}$ to have the full tight binding model where, $f_{p} = 1 + 2  \text{exp}(-i 3 p_{y}/a) + \cos(\sqrt{3} p_{x} a /2)$. Here, $p_{x}$ and $p_{y}$ are the two crysal momentum along $x$ and $y$ directions respectively. 
We also consider the effect of asymmetric gate voltage and SOC on the band structure of uDBLG and depict in Fig.~\ref{uDBLG(AB-AB)_bands}. For a gate voltage of $\Delta=5.2$ meV 
and SOC strength of $\lambda_{so} =1$ meV, we observe that the interplay between intrinsic-SOC and gate-voltage opens up a band-gap ($\approx$ 20 meV) around valley-$K^{\prime}$ 
[see the inset (a) of Fig.~\ref{uDBLG(AB-AB)_bands}] and bands remain gapless at valley-$K$ as highlighted in the top right corner inset (b) of Fig.~\ref{uDBLG(AB-AB)_bands}. Note that, since spin is included in our model, valley-$K$ and valley-$K^{\prime}$ are no longer the time reversal partner of each other, hence we observe the asymmetric band gaps at the following two points. 
In this system the time reversal partners are valley-$K$ for spin up (\ie $K,\uparrow$) with valley-$K^{\prime}$ for spin down (i.e. $K^{\prime},\downarrow$) and valley-$K$ for spin down (i.e. $K,\downarrow$) with  valley-$K^{\prime}$ for spin up (i.e. $K^{\prime},\uparrow$). For the same reason, it is observed that band gap opens up at valley-$K$, and bands are gapless at valley-$K^{\prime}$ (not shown here) for the down spin sector. Such behavior of the bands trigger topological phase transitions in presence of electric field and SOC as we discuss in the next section.

\begin{figure*}[t]
	\centering
	\subfigure{\includegraphics[width=1.0\textwidth]{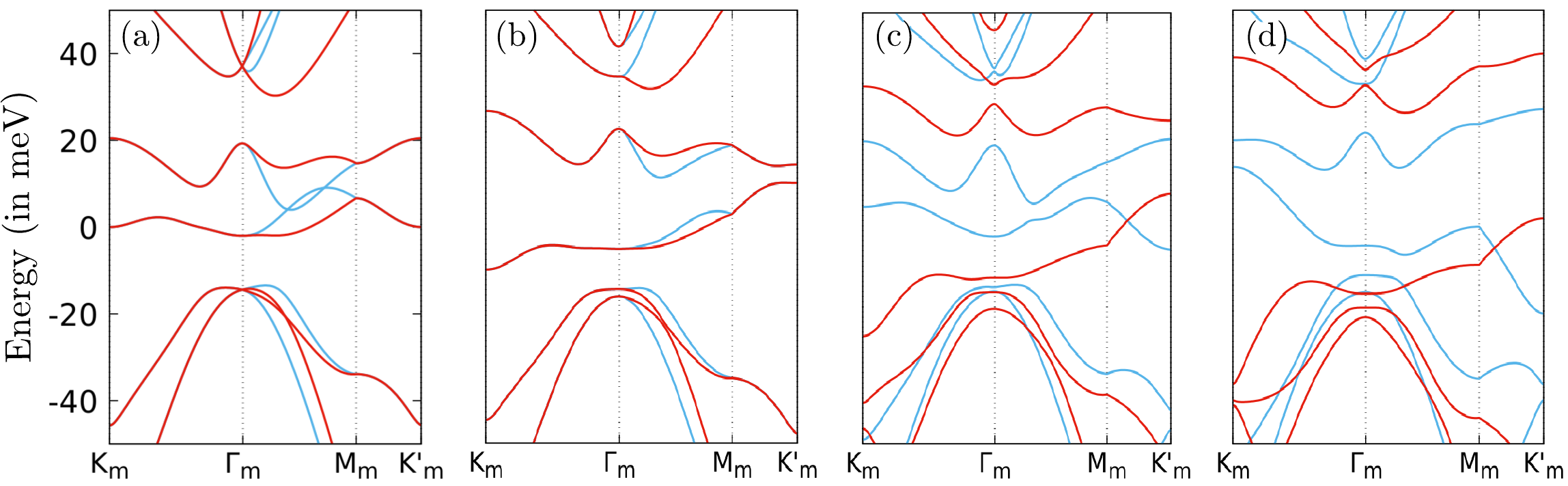}}
	\caption{The electronic band dispersion of the AB-AB tDBLG along the high symmetry path 
is illustrated in presence of different values of gate-voltage ($\Delta$) and SOC ($\lambda_{so}$). 
We choose the twist angle $\theta = 1.30^{o}$. 
In panel (a)~$\Delta$ = 0.0 meV and $\lambda_{so}$ = 0.0 meV, panel (b)~$\Delta$ = 0.0 meV and 
$\lambda_{so}$ = 10.0 meV, panel (c)~$\Delta$ = 10.0 meV and $\lambda_{so}$ = 10.0 meV and panel (d)~$\Delta$ = 20.0 meV and $\lambda_{so}$ = 10.0 meV.  Here, the bands around the valley $K$ 
and $K^{\prime}$ are shown by cyan and red colors respectively. 
	}
	\label{tDBLG(AB-AB)_bands}
\end{figure*}

\subsection{Band spectrum of tDBLG}
For the band structure calculation of the tDBLG, we confine ourselves at very small twist angle 
(here, $\theta=1.30^{o}$) as the low-energy band structure becomes prominent and more physical around such angles. In Fig.~($\ref{tDBLG(AB-AB)_bands}$) we show the electronic band structure 
of the AB-AB stacked tDBLG, around both valley-$K$ and valley-$K^{\prime}$. 
Note that, all the band structure plots are shown considering the spin-up sector ($s_{z}=+1$). 
Since we are interested in the low energy bands, hence we only consider the first three valence 
and conduction bands. The four plots of the band structure are shown for different $\Delta$ and 
$\lambda_{so}$. 
In Fig.~$\ref{tDBLG(AB-AB)_bands}$(a), we observe the band touching between lowest valence and conduction band of $K$-valley bands when $\Delta = \lambda_{so}=0$. However, a gap opens up
in the presence of non-zero $ \lambda_{so}$ as shown in Fig.~$\ref{tDBLG(AB-AB)_bands}$(b). 
Finally, as we turn on both $\Delta$ and $\lambda_{so}$, the band gap increases and the bands 
become more dispersive as we increase the value of them [see Figs.~$\ref{tDBLG(AB-AB)_bands}$(c)-(d)].

\begin{figure*}[t]
	\centering
	\subfigure{\includegraphics[width=1.0\textwidth]{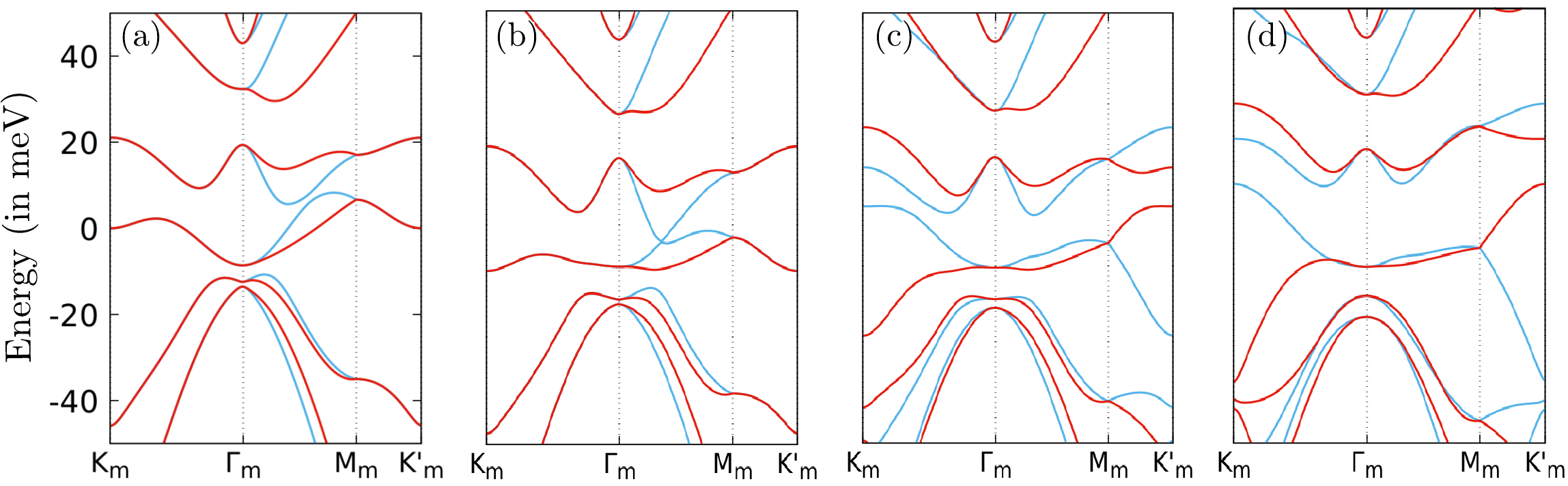}}
	\caption{The electronic band dispersion of the AB-BA stacked tDBLG along the high symmetry path is demonstrated in presence of different values of gate-voltage ($\Delta$) and SOC ($\lambda_{so}$). We choose the twist angle $\theta = 1.30^{o}$. In panels (a)-(d), we choose the same values of
model parameters $\Delta$ and $\lambda_{so}$ as mentioned in Fig.~\ref{tDBLG(AB-AB)_bands}.
Here also, the bands around the valley $K$ and $K^{\prime}$ are shown by cyan and red colors respectively.
	}
	\label{tDBLG(AB-BA)_bands}
\end{figure*}

Similarly, we depict the band structure of AB-BA stacked tDBLG in Fig.~$\ref{tDBLG(AB-BA)_bands}$. Here, Figs.~\ref{tDBLG(AB-BA)_bands}(a)-(d) 
correspond to the band structure for the same set of parameter values as we have considered for 
the AB-AB tDBLG case. At first, we observe that the lowest valence and conduction band (around 
$K$-valley) is gapped in absence of both $\Delta$ and $\lambda_{so}$ [see Fig.~$\ref{tDBLG(AB-BA)_bands}$(a)]. As we switch on only  $\lambda_{so}$ (\ie $\Delta=0$), the lowest two bands become gapless in contrast to the AB-AB tDBLG case as depicted in Fig.~$\ref{tDBLG(AB-BA)_bands}$(b). However, they again become gapped as we turn on both $\Delta$ as well as $\lambda_{so}$ and the 
gap further increases as one enhances the value of gate voltage and SOC. This is shown in 
Fig.~$\ref{tDBLG(AB-BA)_bands}$(c)-(d). Note that, this band gap closing is a feature that arises 
due to the small twist angle $\theta$ and finite $\lambda_{so}$.

\section{Direct band-gap closing and topological phase transitions}\label{Sec:IV}
One of the objectives of this work is to propose topological characterization of the system we have considered. A primary signature towards the topological characterization is to look for  
the band gap closing between the first valence and conduction band. If one observes such closing and reopening of the band gap in the system, one can proceed and perform an appropriate invariant calculation to topologically characterize the system. 

In this section we mainly look for two quantities, one is the direct band gap and the other one is the band gap in the BZ (or in parts of the BZ). The quantitative measurement of the direct band gap 
is done as follows,
\begin{equation}
\delta_{\text{dir}} = \text{min}_{\mathbf{k}}\left[\epsilon_{1}(\mathbf{k}) - \epsilon_{-1}(\mathbf{k}))\right] \in \text{BZ}\ ,
\end{equation}
where, $\epsilon_{-1}(\mathbf{k})$, $\epsilon_{1}(\mathbf{k})$ correspond to the first valence and conduction bands respectively at momentum $\mathbf{k}$. 
First, we proceed to discuss the trivial case of uDBLG and later we show the results for the twisted one. 

\subsection{uDBLG}
\begin{figure}[h]
	\subfigure{\includegraphics[width=0.48\textwidth]{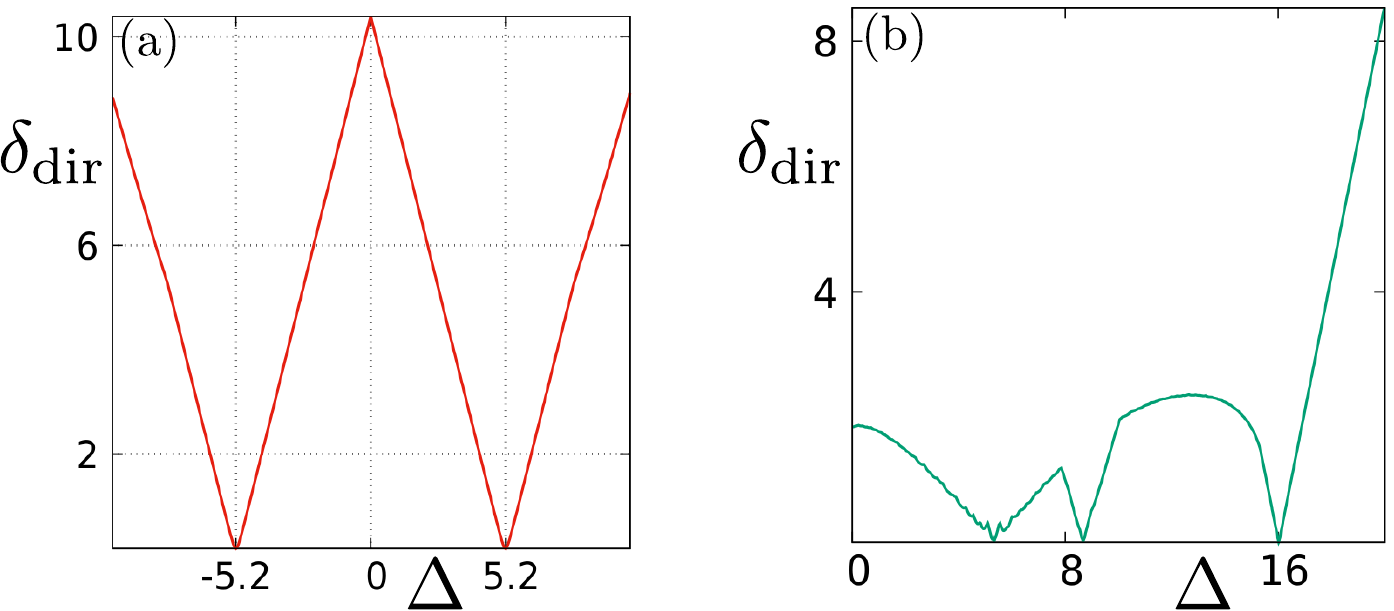}}
	\caption{The behavior of direct-band gap ($\delta_{\text{dir}}$ in meV) is shown as a function of the gate-voltage ($\Delta$ in meV) for (a) uDBLG (choosing $s_{z}=+1$ sector) 
	with $\lambda_{so} = 1.0$ meV 
	and (b) AB-AB tDBLG (at valley-$K$ and $s_{z}=+1$).  We choose $\lambda_{so} = 10.0$ meV and twist angle $\theta = 1.30^{o}$ for panel (b).
	}
	\label{DBLG(AB-AB)_dir_band_gap}
\end{figure}

\begin{figure*}[t]
	\centering
	\subfigure{\includegraphics[width=1.0\textwidth]{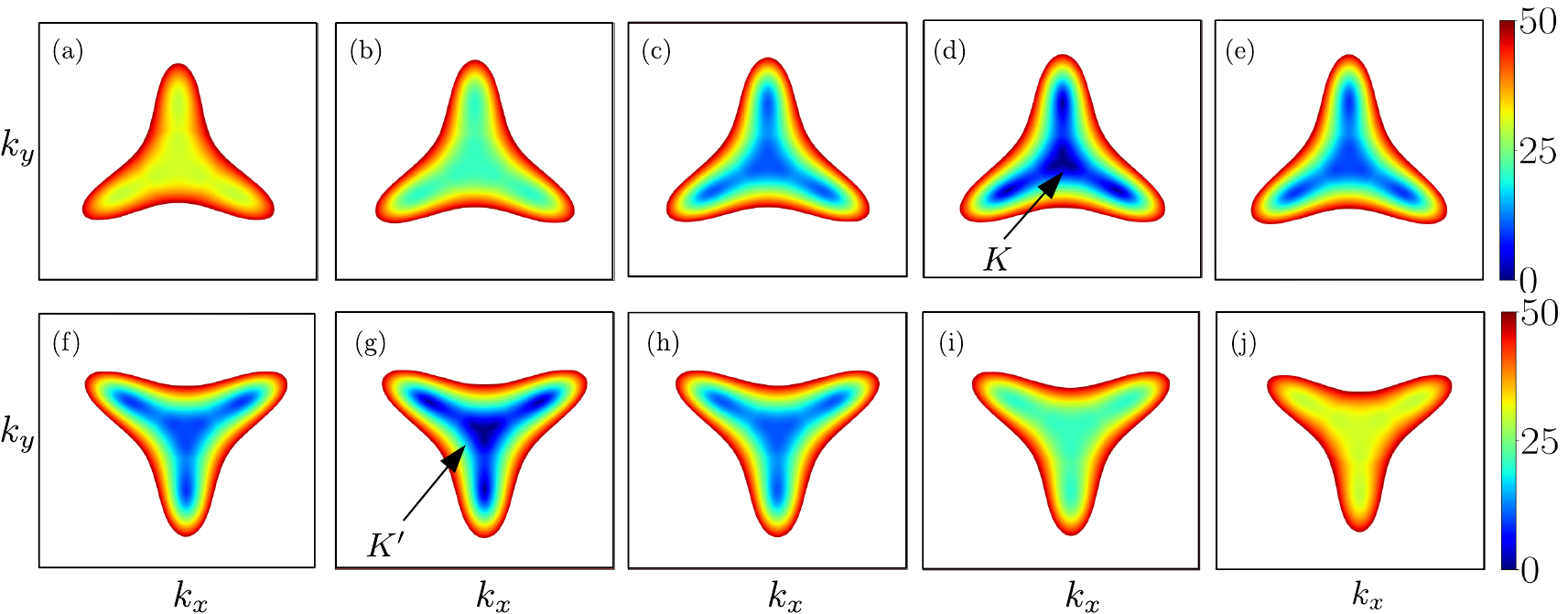}}
	\caption{Density plots for the band gap (in meV) are shown in the $k_{x}$-$k_{y}$ plane within the BZ of uDBLG. In the first column, panel (a)-(e) correspond to the plots for $\Delta = -10.0$ meV, $\Delta = -5.2$ meV, $\Delta = 0.0$ meV, $\Delta = 5.2$ meV, $\Delta = 10.0$ meV respectively near the 
valley-$K$. Whereas, in the second coloumn, panel (f)-(j) indicate the same for the same respective values of the gate-voltages ($\Delta$) near the valley-$K^{\prime}$.
In both the cases, we choose $\lambda_{so} = 1.0$ meV.
	}
	\label{uDBLG(AB-AB)_BZ_band_gap}
\end{figure*}

In Fig.~$\ref{DBLG(AB-AB)_dir_band_gap}$(a), we depict the behavior of direct band gap ($\delta_{\text{dir}}$) as we change the gate voltage ($\Delta$), considering AB-AB uDBLG and $s_{z}=+1$.
It is evident that the direct band gap falls to zero twice for $\Delta = + 5.2$ meV and $\Delta = - 5.2$ meV, symmetrically about $\Delta = 0.0$ meV. Thus, there can be two possible topological phase transition points. 
We also illustrate the density plots for the band gap in the $k_{x}$-$k_{y}$ plane near valley-$K$ and valley-$K^{\prime}$ within the hexagonal BZ in Figs.~$\ref{uDBLG(AB-AB)_BZ_band_gap}$(a)-(j).
 There one can also see how the band gap changes as one tunes the electric field (gate voltage) for a fixed value of SOC. Near each Dirac point, three satellite Dirac points are formed in a triangular shape and the band gap is vanishing at all four points. 
It can also be seen that for $\Delta = 5.2$ meV the band-gap becomes zero at valley-$K$ [see Fig.~\ref{uDBLG(AB-AB)_BZ_band_gap}(d)], while for $\Delta = -5.2$ meV the same is zero at 
valley-$K^{\prime}$ [see Fig.~\ref{uDBLG(AB-AB)_BZ_band_gap}(g)]. This exactly corresponds to the two direct band-gap closings shown in Fig.~$\ref{DBLG(AB-AB)_dir_band_gap}$(a) and they 
indeed take place at two different $\mathbf{k}$-points in the BZ.

\subsection{tDBLG}

For the twisted case, first, we depict the direct band-gap closings in Fig.~\ref{DBLG(AB-AB)_dir_band_gap}(b) for AB-AB stacked tDBLG near valley-$K$ and choosing $s_{z}=+1$. 
Here, we vary the external gate voltage from  $\Delta = 0$ meV to $20$ meV for a fixed value of SOC strength $\lambda_{so} = 10$ meV. We observe that the direct band-gap vanishes at three distinct points [as shown in Fig.~\ref{DBLG(AB-AB)_dir_band_gap}(b)] of the gate voltage. In the absence of SOC, one obtains two direct band-gap closings within the same parameter regime~\cite{Koshino-tDBLG, Mohan2021}. This observation enables us to anticipate the existence of new topological phases in presence of SOC.

\begin{figure*}[t]
	\centering
	\subfigure{\includegraphics[width=1.0\textwidth]{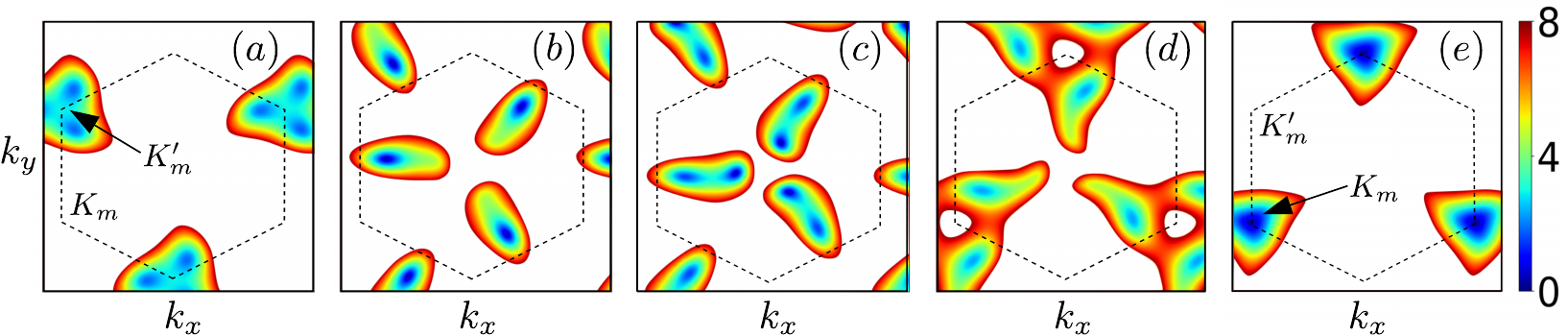}}
	\caption{Density plots for the band gap (in meV) are shown in the $k_{x}$-$k_{y}$ plane within the mBZ 
of AB-AB stacked tDBLG (choosing valley-$K$; $s_{z}=+1$). 
Here panel (a) to (e) correspond to the band gap features for $\Delta = 0.0$ meV, $\Delta = 5.3$ meV, 
$\Delta = 8.7$ meV, $\Delta = 12.0$ meV, and $\Delta = 16.0$ meV respectively. Here, we choose 
$\lambda_{so} = 10.0$ meV and $\theta = 1.30^{o}$. The hexagons refer to mBZ.
	}
	\label{tDBLG(AB-AB)_band_gap_BZ}
\end{figure*}

To gain a better understanding of the gap closings, we investigate the band-gap (between first valence and conduction band) in Fig.~\ref{tDBLG(AB-AB)_band_gap_BZ} considering the mBZ. The dotted hexagonal lines are added to identify the mBZ. For $\Delta = 0.0$ meV, we observe that three satellite Dirac-points are formed near valley-$K_{m}^{\prime}$ as shown in Fig.~\ref{tDBLG(AB-AB)_band_gap_BZ}(a).  
These satellite Dirac points are disassembled from each other with the increasing of gate voltage and 
at the first transition point [see Fig.~\ref{tDBLG(AB-AB)_band_gap_BZ}(b)] the band gap at those 
points vanishes completely for $\Delta = 5.3$ meV. Note that, each satelite Dirac point further splits 
into a rod-like shape and at both ends of that rod shape comparatively smaller gaps are formed. 
At the second transition point [see Fig.~\ref{tDBLG(AB-AB)_band_gap_BZ}(c)], the newly formed gap completely vanishes. Finally, the satellite Dirac points again appear [see Fig.~\ref{tDBLG(AB-AB)_band_gap_BZ}(d)] around the $K_m$ points and if one further enhances $\Delta$, they merge 
to form a central Dirac point at valley-$K_m$ in the mBZ as shown in Fig.~\ref{tDBLG(AB-AB)_band_gap_BZ}(e). There, the band gap again vanishes. In the latter text, we discuss how the 
band-gap closings in the mBZ explain the changes in various Chern numbers while going from one 
topological phase to another.

\section{Topological phase diagrams}\label{Sec:V}
Though the direct-band gap closing indicates the existence of a topological phase transition, nevertheless one needs to investigate an appropriate topological invariant to characterize the phases. 
In case of tDBLG, we calculate the Chern number numerically, employing the discretizion of the mBZ~\cite{Fukui-chern_no}. Since there exists a large number of bands and few of them may overlap 
or cross each other, the appropriate way to proceed in this case is with the non-Abelian Berry connection. Following this we calculate the total Chern number of some $n$-bands, starting from 
$r^{\rm{th}}$ band to $s^{\rm{th}}$ band. 
Hence, we calculate the Chern number as follows,

\begin{eqnarray}
	C_{n} = \frac{1}{2\pi} \int_{mBZ}  d^{2}\mathbf{k}   \hspace{2pt} \text{Tr}({\bf{F}}_{n,\mathbf{k}} \cdot \hat{\mathbf{z}})\ ,
	\label{Eq:Fukui-chern-1}
\end{eqnarray}
where, ${\bf{F}}_{n\mathbf{k}}^{rs}$ is the Berry curvature, integrated over the mBZ and,

\begin{eqnarray}
	{\bf{F}}_{n\mathbf{k}}^{rs} &=& \mathbf{\nabla_{k}} \times {\bf{A}}_{n\mathbf{k}}^{rs} + i [A_{n,kx},A_{n,ky}]^{rs}\ , \\
	{\bf{A}}_{n\mathbf{k}}^{rs} &=& i \langle \psi_{r\mathbf{k}} | \mathbf{\nabla_{k}}|  \psi_{s\mathbf{k}} \rangle\ .
	\label{Eq:Fukui-chern-3}
\end{eqnarray}

Here, ${\bf{A}}_{n\mathbf{k}}^{rs}$ is the corresponding matrix-valued Berry connection. $|\psi_{r\mathbf{k}} \rangle$ are the Bloch states of the Moir\'e superlattice with $r$ being the band index. 
This method works well when the highest band ($s^{\rm{th}}$ band) is separated from its consecutive higher band [$(s+1)^{\rm{th}}$ band] by a direct band gap. 
In our work, for all the topological phase diagrams, the Chern numbers are calculated considering all the filled valence bands into account. Depending upon the Bloch states around which valley and 
what spin sector we are considering, we obtain the corresponding Chern numbers. As we already introduce at the end of Sec.~\ref{Sec:II}, for each block Hamiltonians $H^{\rm{AB-AB}}(K;\uparrow)$, $H^{\rm{AB-AB}}(K;\downarrow)$, $H^{\rm{AB-AB}}(K^{\prime};\uparrow)$, $H^{\rm{AB-AB}}(K^{\prime};\downarrow)$, corresponding Chern numbers can be calculated and represented as 
$C_{\uparrow}^{K}$, $C_{\downarrow}^{K}$, $C_{\uparrow}^{K^{\prime}}$,$C_{\downarrow}^{K^{\prime}}$ respectively. The same prescription for the calculation of Chern numbers follows for the 
AB-BA tDBLG case also.
\subsection{\text{AB-AB}}
\label{SubSec:5A}

\begin{figure}[t]
	\centering
	\subfigure{\includegraphics[width=0.49\textwidth]{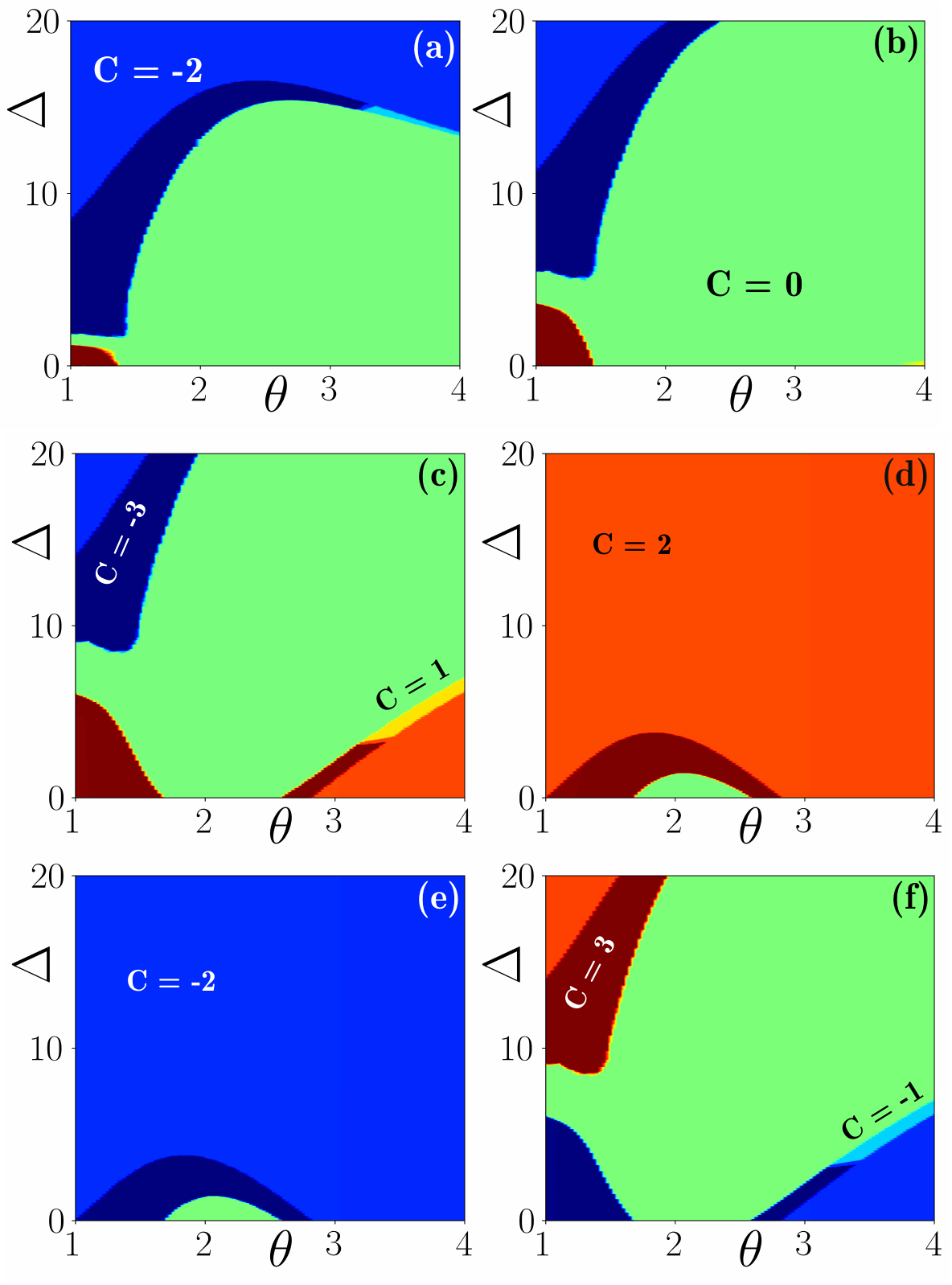}}
	\caption{The density plots for Chern numbers are depicted in the gate voltage ($\Delta$ in meV) and twist angle ($\theta$ in ${o}$) plane for AB-AB tDBLG. In panels (a), (b), and(c) we present the 
	behavior of $C_{\uparrow}^{K}$ with increasing SOC strengths, choosing $\lambda_{so}= 2.0$ meV, $\lambda_{so}= 6.0$ meV and $\lambda_{so}= 10.0$ meV respectively. On the other hand,
	in panels (d), (e), and (f) we showcase the features of $C_{\uparrow}^{K^{\prime}}$, $C_{\downarrow}^{K}$, and $C_{\downarrow}^{K^{\prime}}$ respectively for a fixed SOC strength of 
	$\lambda_{so}= 10.0$ meV. Same colors appearing in different phase diagrams indicate the same values for the corresponding Chern numbers.
	}
	\label{Phase_diag1}
\end{figure}

Here, we depict the topological phase diagrams through several density plots where the Chern numbers ($C_{\uparrow}^{K}$, $C_{\downarrow}^{K}$, $C_{\uparrow}^{K'}$,$C_{\downarrow}^{K'}$) 
are plotted in the gate voltage and twist angle ($\Delta-\theta$) plane, in presence of non-zero SOC strength. In Fig.~\ref{Phase_diag1}(a)-(f), we show 
the effect of increasing SOC strength for $C_{\uparrow}^{K}$. 
In these plots green, dark blue, light blue, maroon, orange, yellow and cyan color regions designate Chern numbers $C = 0$, $C = -3$, $C = -2$, $C = 3$, $C = 2$,  $C = 1$ and $C = -1$ respectively.
The same color regions appearing in any of these six phase diagrams carry the same Chern number values. 
As we increase the SOC strength, we observe that the dark-blue phase (with $C = -3$) region is pushed towards the upside and it becomes narrower. Although this fearure is not prominent for $\lambda_{so} = 2.0$ meV [see Fig.~\ref{Phase_diag1}(a)]. When SOC becomes $\lambda_{so} = 6.0$ meV, it is evident that the phase diagram becomes mostly non-topological ($C = 0$) 
towards the higher twist angles and $\Delta$ as shown in Fig.~\ref{Phase_diag1}(b). 
However, with the further increment in the SOC strength ($\lambda_{so} = 10.0$ meV), we observe that new topological phases ($C = 3$, $C = 2$, $C = 1$) emerge and makes the phase diagram 
[see Fig.~\ref{Phase_diag1}(c)] more interesting. 

\begin{figure}[h]
	\centering
	\subfigure{\includegraphics[width=0.35\textwidth]{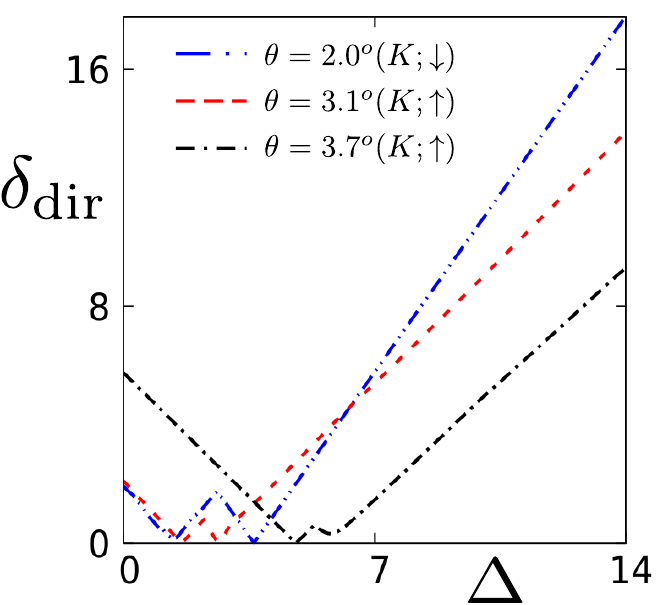}}
	\caption{Direct band gap ($\delta_{\text{dir}}$ in meV) is shown as a function of the gate voltage ($\Delta$ in meV) for AB-AB tDBLG at different twist angles. Here we choose 
	$\lambda_{so}= 10.0$ meV. 
	}
	\label{Direct_band_gap2}
\end{figure}
\begin{figure*}[t]
	\centering
	\subfigure{\includegraphics[width=1.03\textwidth]{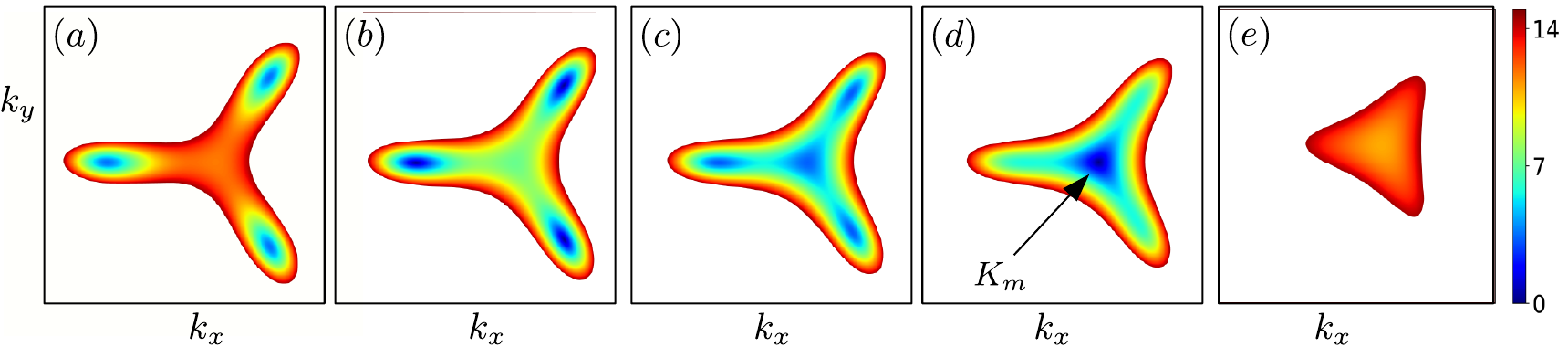}}
	\caption{Density plot for band-gap (in meV) in the mBZ is shown in the $k_{x}$-$k_{y}$ plane for AB-AB tDBLG (valley-$K$; $s_{z}=-1$) choosing a particular twist angle $\theta = 2.0^{o}$. 
	Here, band-gap in panels (a), (b), (c), (d), and (e) are shown near the $K_{m}$ point of the mBZ for $\Delta = 0.0$ meV, $\Delta = 1.4$ meV, $\Delta = 2.6$ meV, $\Delta = 3.63$ meV, and 
	$\Delta = 10.0$ meV respectively. Interestingly, panels (b) and (c) refer to such gap closing points in the mBZ, where in the former the gaps are closing at three satellite Dirac points, 
	and gap closing takes place at the central Dirac point in the latter case. We choose fixed SOC strength $\lambda_{so} = 10.0$ meV. 
	}
	\label{AB-AB_BZ_band_gap-2.0}
\end{figure*}
\begin{figure*}[t]
	\centering
	\subfigure{\includegraphics[width=0.7\textwidth]{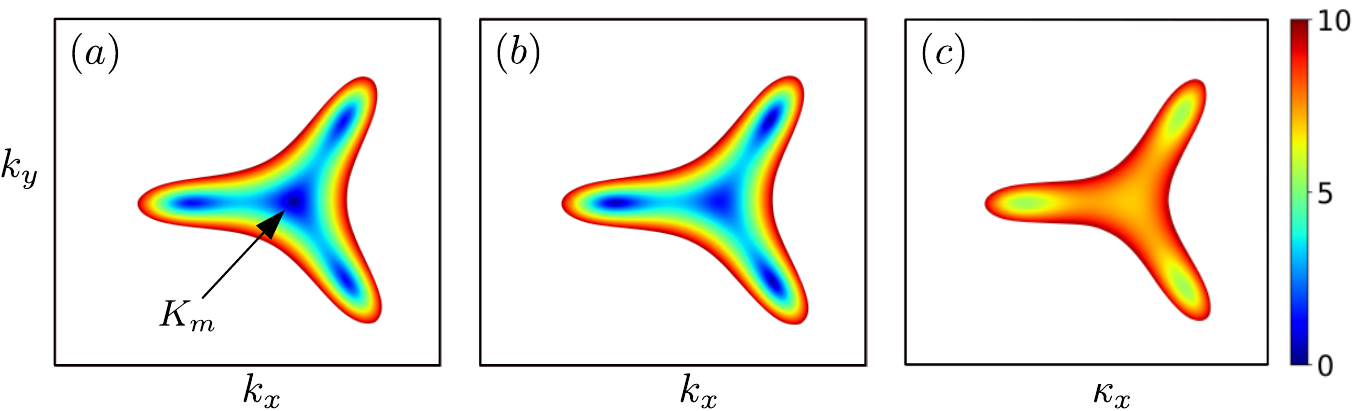}}
	\caption{Density plot for band-gap (in meV) in the mBZ is shown in the $k_{x}$-$k_{y}$ plane for AB-AB tDBLG (valley-$K$; $s_{z}=+1$) at twist angle $\theta = 3.10^{o}$. 
	Here, band-gap in panels (a), (b), and (c) are shown near the $K_{m}$ point of the mBZ choosing $\Delta = 1.62$ meV, $\Delta = 2.67$ meV, and $\Delta = 7.0$ meV respectively. 
	Note that, in panel (a) the gap closes at the central Dirac point of the mBZ, whereas in panel (b) the gap closes at three satellite Dirac points of the same. We choose $\lambda_{so} = 10.0$ meV. 
	}
	\label{AB-AB_BZ_band_gap-3.1}
\end{figure*}

Now we discuss our results for the four topological phase diagrams (depending on the valley and spin degrees of freedoms) of AB-AB stacked tDBLG choosing SOC strength $\lambda_{so} = 10$ meV. 
In Figs.~\ref{Phase_diag1}(c)-(f), we depict the phase diagrams respectively for $C_{\uparrow}^{K}$, $C_{\uparrow}^{K^{\prime}}$, $C_{\downarrow}^{K}$, and $C_{\downarrow}^{K^{\prime}}$. 
As an effect of the SOC, we observe that certain new phases appear in the phase diagrams and some phases seem to disappear from the phase diagrams, depending on the valley and spin sectors. 
In Fig.~\ref{Phase_diag1}(c), for $C_{\uparrow}^{K}$, the phases with $C=1$ (yellow), $C=2$ (orange) and $C=3$ (maroon) are the newly appeared phases as mentioned before. While $C=0$ (green), $C=-2$ (blue), and $C=-3$ (dark blue) are the phases that are present even without any SOC~\cite{Mohan2021}. 
On the other hand, for $C_{\downarrow}^{K}$ the phase with $C=0$ appears in a very small region. Another important observation is that the Chern numbers $C_{\uparrow}^{K}$ [see 
Fig.~\ref{Phase_diag1}(c)] are exactly opposite to the Chern numbers $C_{\downarrow}^{K^{\prime}}$
as shown in Fig.~\ref{Phase_diag1}(f). 
The same is true for $C_{\downarrow}^{K}$ [see Fig.~\ref{Phase_diag1}(e)] and $C_{\uparrow}^{K^{\prime}}$ [see Fig.~\ref{Phase_diag1}(d)]. This behavior of the Chern numbers 
can be attributed to the fact that valley-$K$ is valley-$K^{\prime}$ are the time reversal partner of each other.

Here, we discuss the band gap closings in the mBZ to explain the corresponding change in Chern numbers with the change in chirality around a gap closing point. In Fig.~\ref{Phase_diag1}(c), at 
$\theta = 1.30^{o}$, along the $y$-axis the Chern number changes from $C = +3$ to $C = 0$ when $\Delta = 5.3$ meV. The reason behind this transition are the three satellite Dirac points (around 
$K_{m}^{\prime}$) inside the mBZ and when the valence and conduction band touches each other at the satellite Dirac points (\ie the band-gap vanishes) the chirality index of each changes by  $-1$. 
As a result the total Chern number changes by $3$. In the next transition point, the Chern number changes from $0$ to $-3$ at $\Delta = 8.7$ meV. However, then each satellite Dirac point breaks into a rod-like shape with two smaller gaps at the ends [see Fig.~\ref{tDBLG(AB-AB)_band_gap_BZ}(c)]. Nevertheless, exactly at 
the newly formed point, the valence and conduction band touches each other. These points also exhibit the same chirality index change by $-1$ each. Therefore, we again observe the same change 
in the total Chern number \ie by 3. Finally, at $\Delta = 16.0$ meV the Chern number changes from $-3$ to $-2$. At this scenario, the band-gap closes at the central Dirac point $K_{m}$ [see Fig.~\ref{tDBLG(AB-AB)_band_gap_BZ}(e)] and hence the change in chirality index is $+1$. Thus the Chern number changes by $+1$, and becomes $-2$ from $-3$. 

In Fig.~\ref{Direct_band_gap2} we depict the behavior of direct band gaps as a function of the gate voltage choosing different twist angles. This enables us to understand the topological phase transitions
at other twist angles as portrayed by the phase diagrams in Fig.~\ref{Phase_diag1}. 
In all these line plots of Fig.~\ref{Direct_band_gap2}, the direct band gap closings support a corresponding topological phase transition in the phase diagram at different $\theta$. Further, we also explore the band gap in the mBZ corresponding to the twist angles mentioned in Fig.~\ref{Direct_band_gap2}. In Fig.~\ref{AB-AB_BZ_band_gap-2.0}, we show the band gap features along the line 
$\theta = 2.0^{o}$ for five discrete values of the gate voltages $\Delta$. Following the previous explanation, the change in chirality around a gap-closing point corresponds to the change in Chern number at the topological phase transition point. In this context, one can observe that at $\Delta = 1.4$ meV and $\Delta = 3.63$ meV the change in Chern numbers are respectively by $3$ and $1$. This follows in the phase diagram (at $\theta = 2.0^{o}$) of $C_{\downarrow}^{K}$ where Chern number becomes $-3$ from $0$ at the first transition point and from $-3$ to $-2$ at the second transition point 
[see Fig.~\ref{Phase_diag1}(e)]. Similar gap closing transitions in mBZ are observed along the line $\theta = 3.1^{o}$ as shown in Fig.~\ref{AB-AB_BZ_band_gap-3.1}. 
Here, the gap closing points take place at $\Delta = 1.62$ meV (one gap closing point at the central Dirac point) and at $\Delta = 2.67$ meV (three gap closing points at the satellite Dirac points).
Thus it explains the change in Chern numbers from $2$ to $3$ at the first tansition point and from $3$ to $0$ at the second transition point for $C_{\uparrow}^{K}$ at $\theta = 3.1^{o}$ [see Fig.~\ref{Phase_diag1}(c)]. We have the similar plots for the direct mBZ band gap corresponding to $C_{\uparrow}^{K^{\prime}}$ and $C_{\downarrow}^{K^{\prime}}$ phase diagrams as well.
Nevertheless, as these are symmetric to the previously shown phase diagrams, we do not repeat those results here.


Finally, we conclude this sub-section by computing different Chern numbers to characterize our system. We define, the valley Chern number ($C_{v}$),  the spin Chern number ($C_{s}$), the spin-valley Chern number ($C_{sv}$), and the total Chern number ($C_{t}$) for the tDBLG, to convey a proper topological characterization of the system. Below we present the expressions~\cite{Ezawa_2012-Diff_chern} for calculating the above mentioned Chern numbers as,
\begin{eqnarray}
\label{diff-chern-definitions_c_v}
C_{v} &=& C_{\uparrow}^{K} - C_{\uparrow}^{K'} + C_{\downarrow}^{K} - C_{\downarrow}^{K'}\ , \\
\label{diff-chern-definitions_c_s}
C_{s} &=& \frac{1}{2} \left( C_{\uparrow}^{K} + C_{\uparrow}^{K'} - C_{\downarrow}^{K} - C_{\downarrow}^{K'} \right)\ , \\
\label{diff-chern-definitions_c_sv}
C_{sv} &=& \frac{1}{2} \left( C_{\uparrow}^{K} - C_{\uparrow}^{K'} - C_{\downarrow}^{K} + C_{\downarrow}^{K'} \right)\ , \\
\label{diff-chern-definitions_c_tot}
C_{t} &=& C_{\uparrow}^{K} + C_{\uparrow}^{K'} + C_{\downarrow}^{K} + C_{\downarrow}^{K'}\ ,
\end{eqnarray}

\begin{figure}[H]
	\centering
	\subfigure{\includegraphics[width=0.5\textwidth]{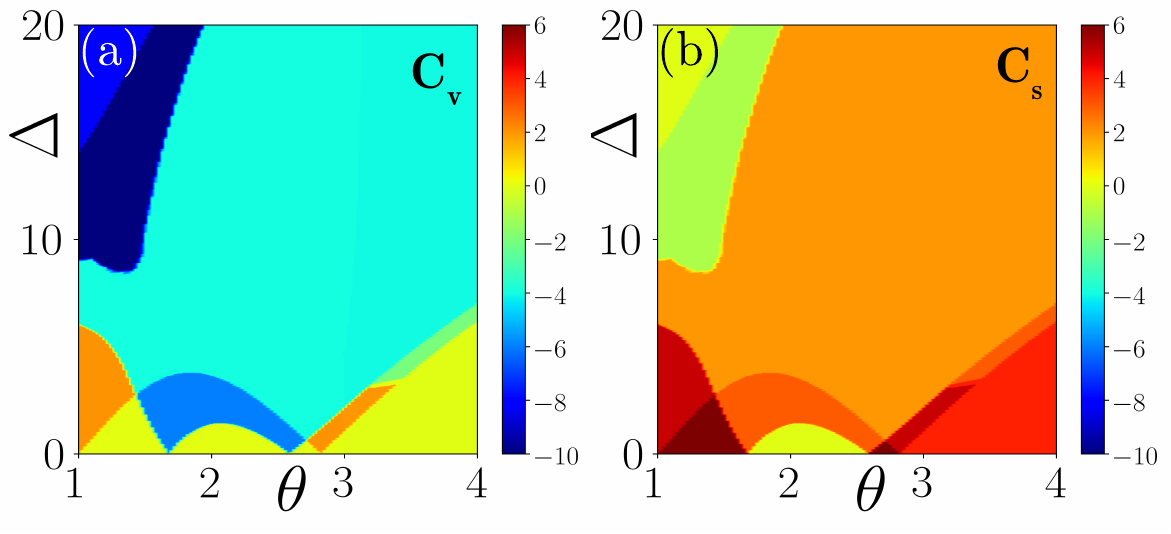}}
	\caption{The topological phase diagram corresponding to the (a) valley Chern number ($C_{v}$), (b) spin-Chern number ($C_{s}$) is illustrated for the AB-AB tDBLG in the $\Delta - \theta$ plane.
	Here, we choose $\lambda_{so} = 10.0$ meV.	
	}
	\label{Phase_diag_diff_ch_no-AB-AB}
\end{figure}

Since in our low energy continuum model analysis, we neither consider intervalley mixing (in absence of impurities), nor spin-flip scattering (as intrinsic SOC is considered only), thus calculation 
of the above Chern numbers are well defined. Below we discuss the corresponding topological phase diagrams for AB-AB tDBLG. See latter text for AB-BA tDBLG discussion. 
In Fig.~\ref{Phase_diag_diff_ch_no-AB-AB} we show the phase diagrams corresponding to the above-defined Chern numbers for the AB-AB tDBLG. In the two phase diagrams [\ie Fig.~\ref{Phase_diag_diff_ch_no-AB-AB}(a) and Fig.~\ref{Phase_diag_diff_ch_no-AB-AB}(b)], the valley Chern number ($C_{v}$) and the spin Chern number ($C_{s}$) are shown in the plane of gate-voltage and 
twist angle (\ie $\Delta - \theta$). Note that, the spin-valley Chern number ($C_{sv}$) and the total Chern number ($C_{t}$) vanish throughout the $\Delta - \theta$ parameter space. The different values 
of the Chern numbers corresponding to different regions of the parameters space are indicated by the color bar for each phase diagram. From Figs.~\ref{Phase_diag_diff_ch_no-AB-AB}(a)-(b) it is evident that the AB-AB tDBLG behaves as a quantum valley Hall insulator (QVHI), as well as a quantum spin Hall insulator (QSHI) in most of the parameter regime.  
The only regions with yellow color indicate the zero Chern number \ie trivial regions. 
Thus the presence of SOC turns the AB-AB tDBLG into a QSHI at small twist angles. This is one of the main results of our analysis. From careful observation, we note that there are regions in the 
parameter space where the system is QVHI but not a QSHI and vice versa. Moreover, there are regions where the system becomes both QVHI and QSHI. 
The reason for the total Chern number to be zero is that, while the low energy continuum model that is written around a valley of the original graphene BZ exhibits broken time-reversal symmetry, but 
the total system (including two valleys) preserves the time-reversal symmetry.

\subsection{\text{AB-BA}}\label{SubSec:6A}\label{SubSubSec:6A1}

\begin{figure}[H]
	\centering
	\subfigure{\includegraphics[width=0.48\textwidth]{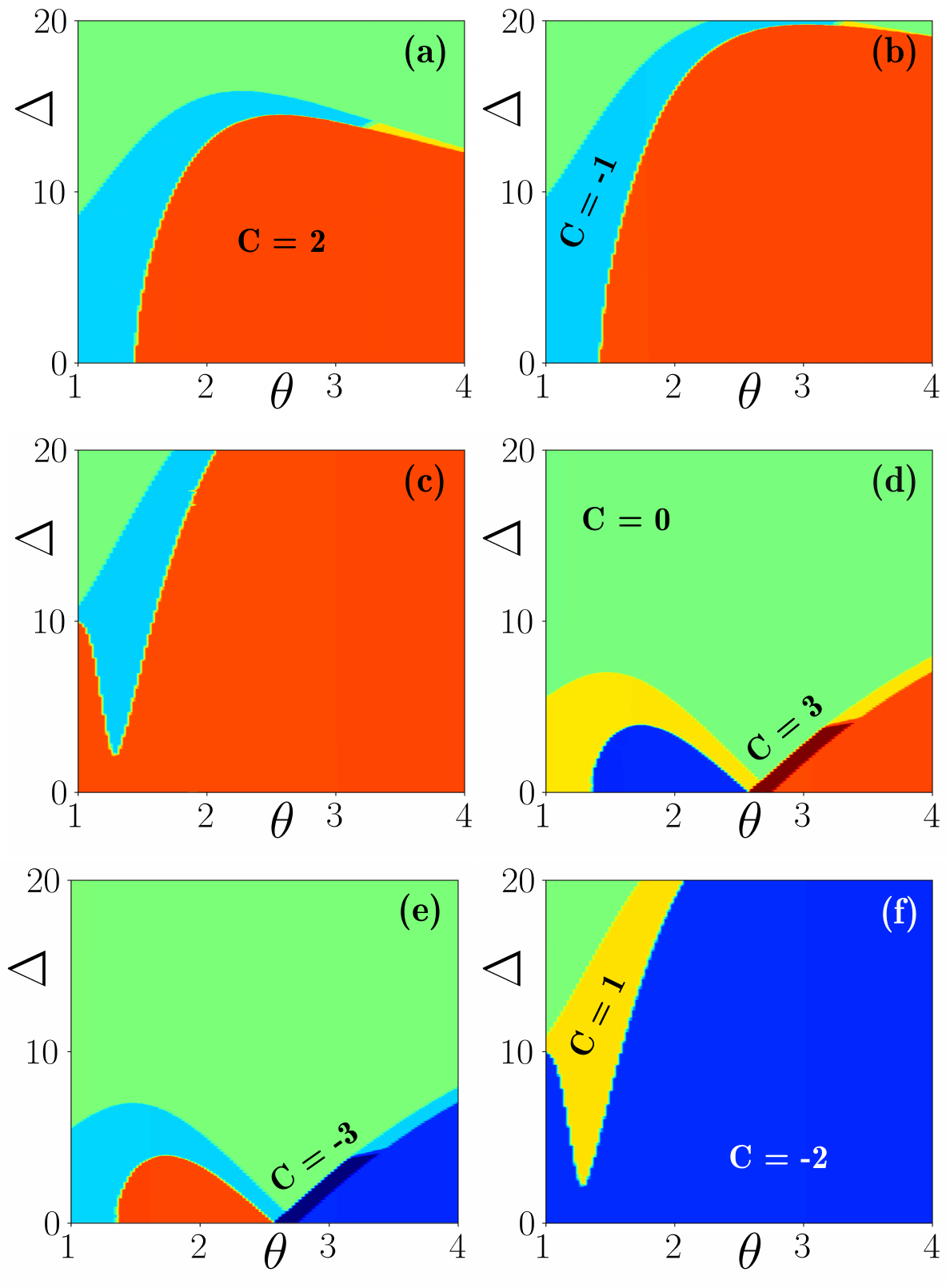}}
	\caption{The density plots for Chern numbers in the gate voltage ($\Delta$ in meV) and twist angle plane for AB-BA tDBLG are shown. In panels (a), (b), and (c) we show the features 
	of $C_{\uparrow}^{K}$ with increasing SOC strengths, choosing $\lambda_{so}= 2.0$ meV, $\lambda_{so}= 6.0$ meV, and $\lambda_{so}= 10.0$ meV respectively. On the other hand, in panels (d), (e), 
	and (f) we depict $C_{\uparrow}^{K^{\prime}}$, $C_{\downarrow}^{K}$, and $C_{\downarrow}^{K^{\prime}}$ for a fixed SOC strength $\lambda_{so}= 10.0$ meV. Note that, same colors appearing 
	in the different phase diagram indicate the same values for the Chern numbers.
	}
	\label{AB-BA_Phase_diag}
\end{figure}

Similar to AB-AB tDBLG, we also investigate the AB-BA case and discuss the effect of increasing 
SOC strength on the topological phase diagram depicted in Fig.~\ref{AB-BA_Phase_diag}. 
In particular, Figs.~\ref{AB-BA_Phase_diag}(a)-(c) correspond to the topological phase diagram of
$C_{\uparrow}^{K}$ for $\lambda_{so} = 2.0$ meV, $\lambda_{so} = 6.0$ meV, and $\lambda_{so} = 10.0$ meV respectively. 
In all the density plots of Fig.~\ref{AB-BA_Phase_diag}, yellow, orange, maroon, green, sky blue, blue 
and navy blue colors represent phases with Chern numbers $C = 1$, $C = 2$, $C = 3$, $C = 0$, 
$C = -1$, $C = -2$ and $C = -3$ respectively. 
Note that, in AB-BA tDBLG case, topological regime with different Chern numbers (\ie $C_{\uparrow}^{K}$ $\neq$ 0) dominates over the trivial regime (\ie $C_{\uparrow}^{K}$ = 0) as one increases the SOC strength. This is in contrast to the AB-AB tDBLG case where major part of the phase diagram remains 
non-topological (\ie $C_{\uparrow}^{K}$ = 0) even with non-zero $\lambda_{so}$.

On the other hand, in Figs.~\ref{AB-BA_Phase_diag}(d)-(f), 
we showcase the density plots of 
$C_{\uparrow}^{K^{\prime}}$, $C_{\downarrow}^{K}$, $C_{\downarrow}^{K^{\prime}}$ respectively in the $\Delta$-$\theta$ plane for a fixed SOC strength of $\lambda_{so} = 10.0$ meV.
Although the values of the Chern numbers and regions of the corresponding phases are very different compared to the AB-AB tDBLG, the overall feature remains the same here too. Here also we observe that the phase diagrams corresponding to $C_{\uparrow}^{K}$ [see Fig.~\ref{AB-BA_Phase_diag}(c)] and $C_{\downarrow}^{K^{\prime}}$ [see Fig.~\ref{AB-BA_Phase_diag}(f)] exhibit exactly opposite 
Chern numbers for each values of the model parameters (\ie $\Delta$ and $\theta$). This remains true for $C_{\uparrow}^{K^{\prime}}$ and $C_{\downarrow}^{K}$ as shown in 
Fig.~\ref{AB-BA_Phase_diag}(d) and Fig.~\ref{AB-BA_Phase_diag}(e) respectively. 
We also observe some new phases appear [$C = -2$ and $C = -3$ in Fig.~\ref{AB-BA_Phase_diag}(e), $C = 2$ and $C = 3$ in Fig.~\ref{AB-BA_Phase_diag}(d)] in these phase diagrams 
which is the consequence of the finite SOC. 

\begin{figure}[H]
	\centering
	\subfigure{\includegraphics[width=0.4\textwidth]{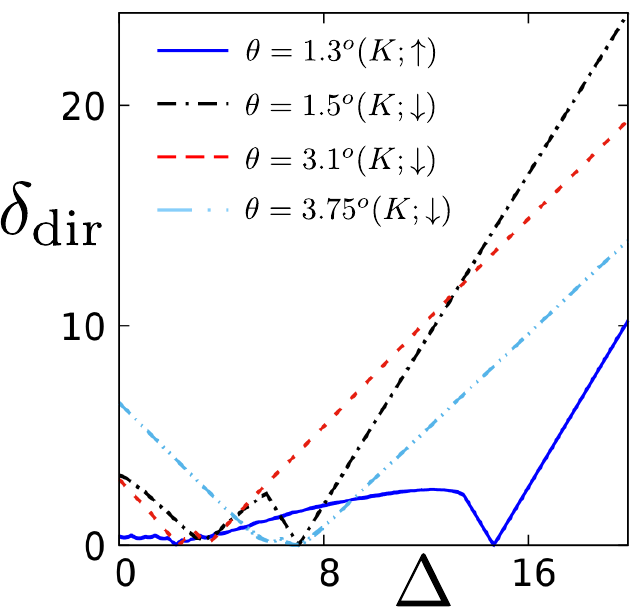}}
	\caption{Direct band gap ($\delta_{\text{dir}}$ in meV) is shown as a function of the gate voltage ($\Delta$ in meV) choosing $\lambda_{so}= 10.0$ meV. Here, $\theta=1.30^{o}$ denotes the 
	band gap closing for valley-$K$; $s_{z}=+1$ sector. On the other hand, $\theta=1.50^{o}$, $\theta=3.10^{o}$, and $\theta=3.75^{o}$ correspond to the same for valley-$K$; $s_{z}=-1$ sector
	of AB-BA tDBLG.
	}
	\label{AB-BA_Direct_band_gap3}
\end{figure}

\begin{figure*}[t]
	\centering
	\subfigure{\includegraphics[width=1.0\textwidth]{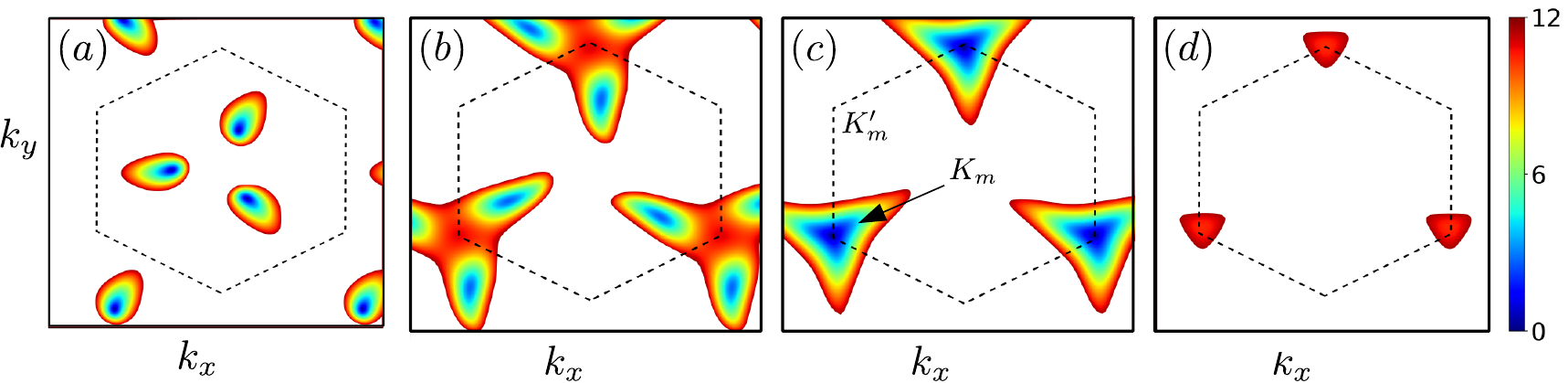}}
	\caption{Density plot for the band-gap (in meV) is shown in the mBZ ($k_{x}$-$k_{y}$ plane) considering valley-$K$; $s_{z}=+1$. 
         Here panels (a) to (d) are depicted for $\Delta = 2.21$ meV, $\Delta = 10.0$ meV, $\Delta = 14.65$ meV, and $\Delta = 20.0$ meV respectively. 
	We choose the twist angle $\theta = 1.30^{o}$ and SOC strength $\lambda_{so} = 10.0$ meV. 
	}
	\label{AB-BA_BZ_band_gap2}
\end{figure*}

\begin{figure*}[t]
	\centering
	\subfigure{\includegraphics[width=0.7\textwidth]{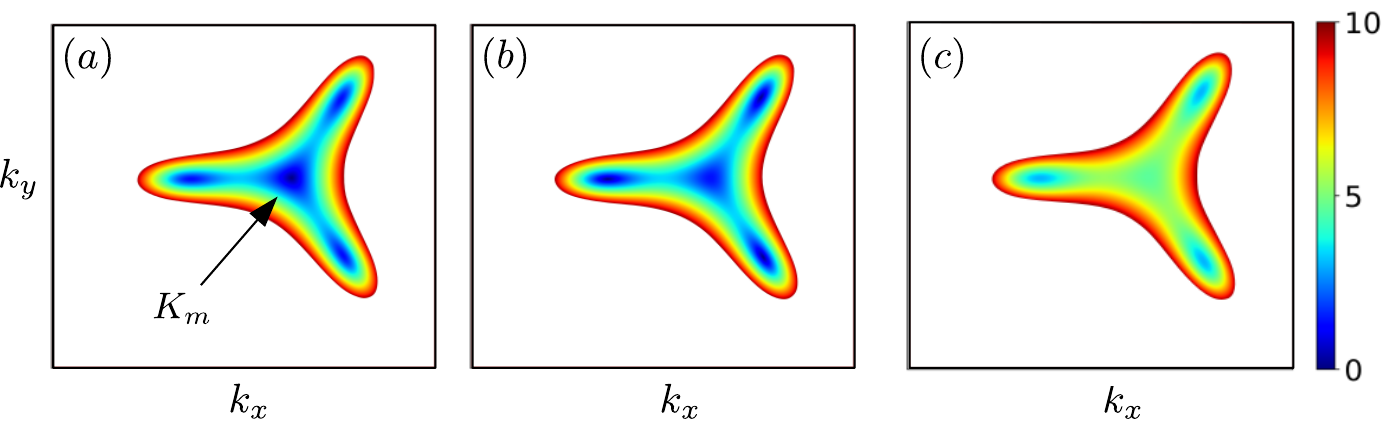}}
	\caption{Density plot for the band-gap (in meV) is shown in the mBZ ($k_{x}$-$k_{y}$ plane) considering valley-$K$; $s_{z}=-1$. Here, band gaps in panels (a), (b),and (c) are depicted near the 
	$K_{m}$ point of the mBZ choosing $\Delta = 2.33$ meV, $\Delta = 3.44$ meV, and $\Delta = 6.0$ meV respectively. Note that, in panel (a) the gap closes at the central Dirac point of the mBZ and in 
	panel (b) the gap closes at three satellite Dirac points of the mBZ. The other model parameters are chosen as $\theta = 3.10^{o}$, $\lambda_{so} = 10.0$ meV. 
	}
	\label{AB-BA_BZ_band_gap3}
\end{figure*}

In order to validate the newly appeared phases, we also compute the direct band gap (shown in Fig.~\ref{AB-BA_Direct_band_gap3}) and band gap in the mBZ (see Fig.~\ref{AB-BA_BZ_band_gap2} 
and Fig.~\ref{AB-BA_BZ_band_gap3}) along few topological phase transition points. In Fig.~\ref{AB-BA_Direct_band_gap3}, the blue line corresponds to the topological phase transitions in the  $C_{\uparrow}^{K}$ phase diagram at $\theta = 1.30^{o}$. On the other hand,
the dashed black, red and the cyan (dashed-dot) color lines indicate the topological phase transitions in the $C_{\downarrow}^{K}$ phase diagram at $\theta = 1.50^{o}$, $\theta = 3.10^{o}$, 
and $\theta = 3.75^{o}$ respectively. 
Note that, at each topological phase transition, the direct band gap falls to zero. Finally to understand the changes in the Chern number, as the AB-AB tDBLG case, we depict the band gap in the mBZ 
along a few of the above direct band gap lines. For \eg in Figs.~\ref{AB-BA_BZ_band_gap2}(a)-(d), 
band gaps in mBZ are shown respectively for $\Delta = 2.21$ meV, $\Delta = 10.0$ meV, $\Delta = 14.65$ meV, and $\Delta = 20.0$ meV along the $\theta = 1.3^{o}$ line in the $C_{\uparrow}^{K}$ phase diagram. Here, Figs.~\ref{AB-BA_BZ_band_gap2}(a) and (c) designate the band gap closings and change in Chern numbers by $3$ and $1$ respectively, which is consistent with the topological phase transition points (note that in Fig.~\ref{AB-BA_Phase_diag}(c), along $\theta = 1.30^{o}$ line, at the first topological phase transition point Chern number changes from $C = 2$ to $C = -1$ and at the 
second transition point it becomes $C = 0$ from $C = -1$). Similarly, we show the mBZ band gap features along the $\theta = 3.10^{o}$ line (see Fig.~\ref{AB-BA_BZ_band_gap3}) corresponding to the $C_{\downarrow}^{K}$ phase diagram [see Fig.~\ref{AB-BA_Phase_diag}(e)]. Therefore, we discuss the direct band gap plots and their connection with the topological phase transitions for 
$C_{\uparrow}^{K}$ and $C_{\downarrow}^{K}$. However, we don't repeat the same for $C_{\uparrow}^{K^{\prime}}$ and $C_{\downarrow}^{K^{\prime}}$ as these phase diagrams are symmetric 
to the previously shown phase diagrams.

\begin{figure}[H]
	\centering
	\subfigure{\includegraphics[width=0.49\textwidth]{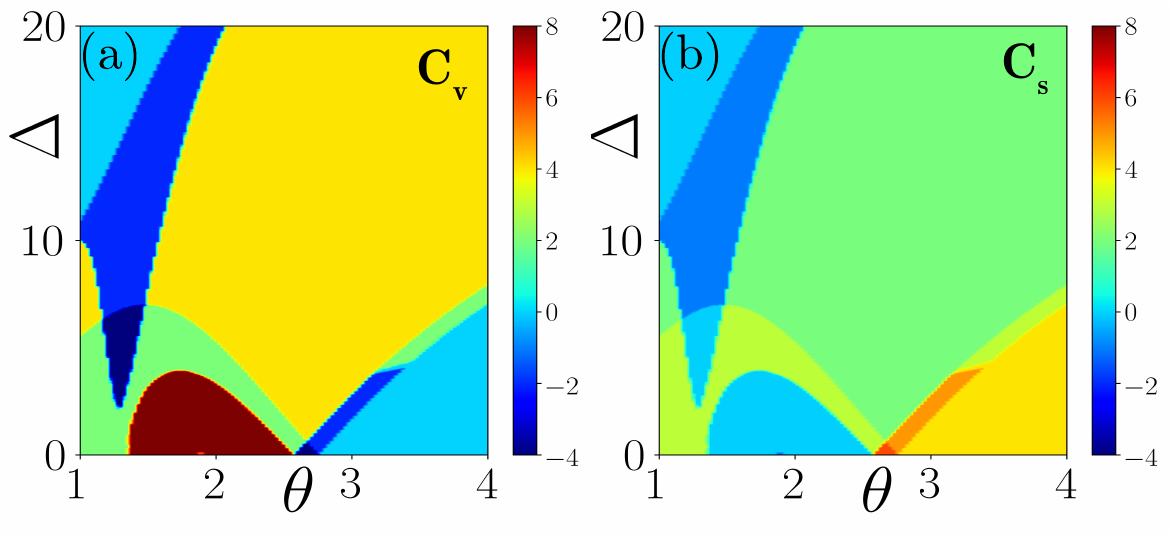}}
	\caption{The topological phase diagram corresponding to the (a) the valley Chern number ($C_{v}$), (b) spin Chern number ($C_{s}$) is illustrated for AB-BA tDBLG in the $\Delta - \theta$ plane. 
	Here, we choose $\lambda_{so}=10.0$ meV. 	
	}
	\label{Phase_diag_diff_ch_no-AB-BA}
\end{figure}

Similar as before, we conclude this subsection by calculating different Chern numbers to characterize the system. We employ the same definitions as mentioned in Eq.~(\ref{diff-chern-definitions_c_v}), 
Eq.~(\ref{diff-chern-definitions_c_s}), Eq.~(\ref{diff-chern-definitions_c_sv}), and  Eq.~(\ref{diff-chern-definitions_c_tot}) to compute respectively the valley Chern number ($C_{v}$), the spin Chern number ($C_{s}$), the spin-valley Chern number ($C_{sv}$), and the total Chern number ($C_{t}$). In Figs.~\ref{Phase_diag_diff_ch_no-AB-BA}(a)-(b), we show the valley Chern numbers ($C_{v}$) and the spin Chern numbers ($C_{s}$) in the gate voltage($\Delta$) and twist angle ($\theta$) plane for a fixed SOC strength of $\lambda_{so} = 10.0$ meV. The total Chern numbers ($C_{t}$) and the spin-valley 
Chern numbers ($C_{sv}$) remain zero everywhere in the $\Delta$-$\theta$ plane, as the entire system (including two valleys) respects time-reversal symmetry. 
From the above two phase diagrams, it is clear that there are regions in the parameter space where it is neither a QVHI nor a QSHI (top left corner of the phase diagrams). Rather the system is 
non-topological in that region with $C = 0$. Then there are regions where it is either a QVHI [see Fig.~\ref{Phase_diag_diff_ch_no-AB-BA}(a)] or QSHI phase emerges [see Fig.~\ref{Phase_diag_diff_ch_no-AB-BA}(b)]. In the rest of the parameter space, the system exhibits both QSHI and QVHI phase. In the absence of the SOC (\ie when only the gate voltage is present), the system is a QVHI~\cite{Koshino-tDBLG, Chebrolu, Mohan2021}. Thus, intrinsic SOC turns both the AB-AB and AB-BA tDBLG into a QSHI where the system is otherwise trivial even in the presence of $\Delta$. Also, the twist angle 
provides a tunability to access different topological phases (QVHI and QSHI). These are also the major findings of our analysis.

\subsection{Comparison between the twisted and untwisted double-bilayer graphene}\label{SubSec:5C}
\begin{figure*}[t]
	\centering
	\subfigure{\includegraphics[width=0.95\textwidth]{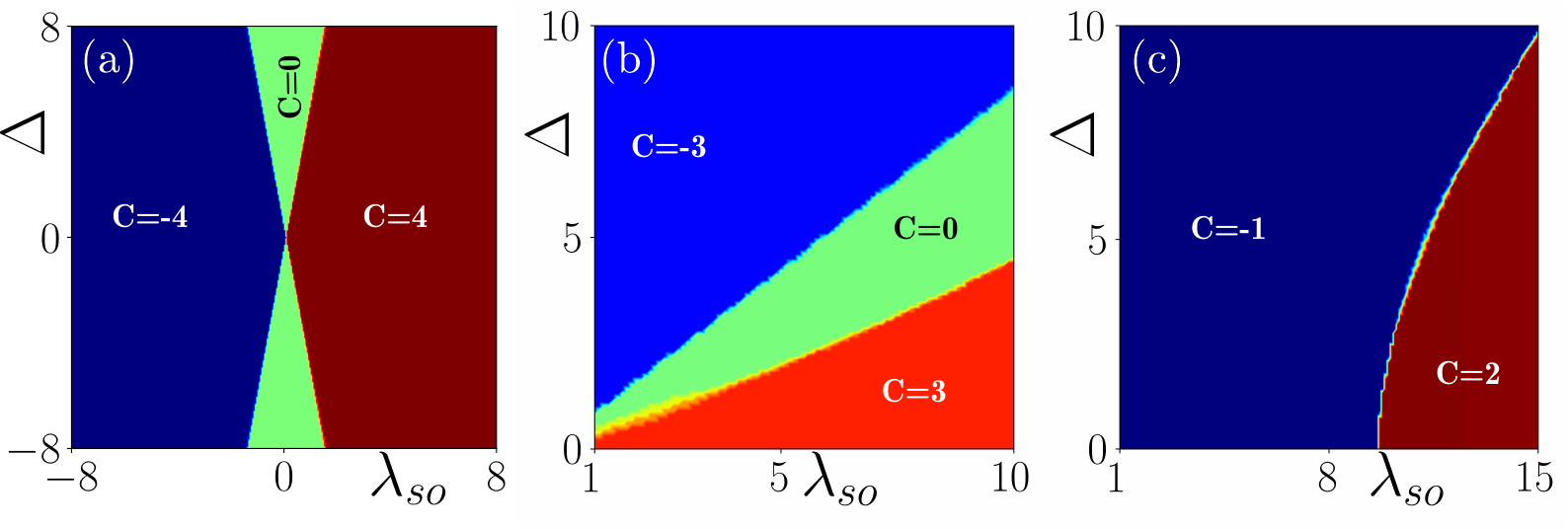}}
	\caption{Density plots for the Chern numbers are depicted in the gate voltage ($\Delta$ in meV) and SOC strength ($\lambda_{so}$ in meV) plane. Here, panels (a), (b) and (c) refer to the 
	Chern number $C_{\uparrow}^{K}$ for AB-AB uDBLG, AB-AB tDBLG, and AB-BA tDBLG respectively. We choose the twist angle $\theta = 1.3^{o}$ for panels (b) and (c). 
	}
	\label{untwisted_chern_phase_diag1}
\end{figure*}

Here, we present a comparative analysis between the topological properties of untwisted and twisted DBLG. In particular, we analyse the Chern number in the gate voltage and SOC strength ($\Delta$-$\lambda_{so}$) plane considering both the systems. In Fig.~\ref{untwisted_chern_phase_diag1}(a) we depict the Chern number for the AB-AB uDBLG in the $\Delta$-$\lambda_{so}$ plane considering the $s_{z}=+1$ sector. On the other hand, in Fig.~\ref{untwisted_chern_phase_diag1}(b) and Fig.~\ref{untwisted_chern_phase_diag1}(c) we illustrate the Chern numbers respectively for AB-AB and AB-BA tDBLGs at a fixed twist angle of $\theta = 1.30^{o}$ considering only valley-$K$ and $s_{z}=+1$. In case of the AB-AB uDBLG, we observe that the topological phase diagram mimics that of the Kane-Mele 
model, with a variation in the values of the Chern numbers ($C=-4$ $C=0$ and $C=4$) that 
arises due to the coupled four layers of uDBLG in presence of SOC. This is consistent with two symmetric gap closing points [see Fig.~\ref{DBLG(AB-AB)_dir_band_gap}(a)] as one varies $\Delta$ at
$\lambda_{so} = 1.0$ meV. The corresponding Chern number changes from $C=0$ to $C=4$ and again $C=4$ to $C=0$. 
In contrast, in the presence of a twist in the system, the phase diagram turns out to be quite different. Here, not only the values of the Chern numbers but also the regions of the topological phases are different than the untwisted case. It is evident from Fig.~\ref{untwisted_chern_phase_diag1}(b) that one can have a topological phase with $C=3$ even if $\Delta=0$. Also, at $\lambda_{so} = 1.0$ meV,
one can have three topological phases ($C=3$, $C=0$, $C=-3$) along the $\Delta$ line compared to two in the uDBLG case. 
Furthermore, for the AB-BA tDBLG, we possess the topological phase $C=-1$ when $\Delta=0$ and single topological phase transition from $C=-1$ to $C=2$. 
This is illustrated in Fig.~\ref{untwisted_chern_phase_diag1}(c). 
Hence, we infer that presence of a twist and SOC changes the topological phase diagrams in case of both AB-AB and AB-BA stacked tDBLG.
\section{Observation of unusual band-gap closings in mBZ}\label{Sec:VI}
\begin{figure*}[t]
	\centering
	\subfigure{\includegraphics[width=1.0\textwidth]{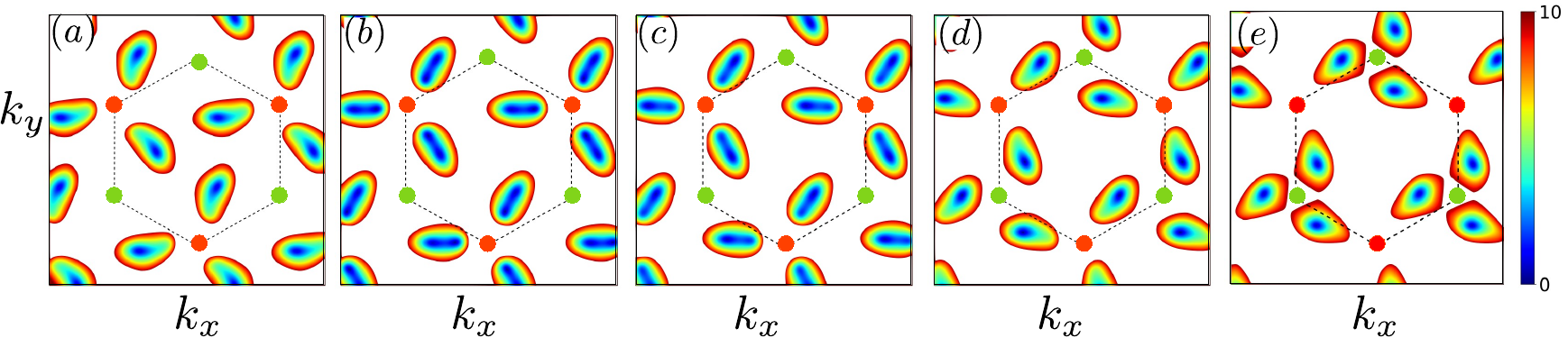}}
	\caption{Band-gap (between the first valence and conduction bands) features (in meV) in the mBZ is shown in the $k_{x}$-$k_{y}$ plane considering AB-AB tDBLG (valley-$K$: $s_{z}=+1$). 
         Here, panels (a) to (e) correspond to the SOC strengths $\lambda_{so} = 0.0$ meV, $\lambda_{so} =3.5$ meV, $\lambda_{so} = 5.0$ meV, $\lambda_{so} = 7.9$ meV, 
         and $\lambda_{so} = 10.0$ meV respectively. We choose $\Delta = 3.0$ meV and $\theta = 1.30^{o}$.
	}
	\label{tDBLG(AB-AB)_Ez-Lso_band_gap_BZ}
\end{figure*}

We observe that most of the topological phase transitions that we encounter in our analysis correspond to a Lifshitz-like transition. However, there are a few of these which do not refer to the Lifshitz-like transitions for \eg the bands gap closings at a central Dirac point or at a satellite Dirac point. Here we show such a band gap closing in the mBZ, 
considering AB-AB tDBLG at fixed twist angle $\theta=1.30^{o}$. However, we vary the SOC strength keeping the value of the gate voltage fixed at $\Delta = 3.0$ meV.
Afterwards, in Fig.~\ref{untwisted_chern_phase_diag1}(b), imagine a horizontal line along $\Delta = 3.0$ meV. Along this line if we translate from the left to right, one can realize the corresponding 
change in Chern number with the variation of SOC strength $\lambda_{so}$. Correspondingly, in 
Figs.~\ref{tDBLG(AB-AB)_Ez-Lso_band_gap_BZ}(a)-(e), we show the band gap in the mBZ choosing
various values of $\lambda_{so}$ for the valley-$K$; $s_{z}=+1$.   
At the first topological phase transition point [see Fig.~\ref{untwisted_chern_phase_diag1}(b) and Fig.~\ref{tDBLG(AB-AB)_Ez-Lso_band_gap_BZ}(b)] Chern number changes from $-3$ to $0$, 
and at the second transition point [as shown in Fig.~\ref{untwisted_chern_phase_diag1}(b) and also see Fig.~\ref{tDBLG(AB-AB)_Ez-Lso_band_gap_BZ}(d)] from $0$ to $+3$. While there are three satellite Dirac points as can be seen in Fig.~\ref{tDBLG(AB-AB)_Ez-Lso_band_gap_BZ}(a), near the topological phase transitions we observe that those points further split into two different pockets 
(at both ends of a rod-like shape) as depicted in Fig.~\ref{tDBLG(AB-AB)_Ez-Lso_band_gap_BZ}(b) and Fig.~\ref{tDBLG(AB-AB)_Ez-Lso_band_gap_BZ}(c).
However, at the transition point, band gap doesn't close at both the pockets, rather they close at the newly formed rod like structures. This corresponds to the first topological phase transition 
we observe at $\lambda_{so} = 3.5$ meV [see Fig.~\ref{untwisted_chern_phase_diag1}(b)]. The second transition takes place due to the gap closings at the satellite Dirac points formed at 
$\lambda_{so} = 7.9$ meV as shown in Fig.~\ref{tDBLG(AB-AB)_Ez-Lso_band_gap_BZ}(d). Throughout this entire process of varying $\lambda_{so}$, the satellite dirac points which are initially 
around the valley-$K_{m}^{^{\prime}}$ [see near the red dots of Fig.~\ref{tDBLG(AB-AB)_Ez-Lso_band_gap_BZ}(a)] finally end up around valley-$K_{m}$ [see near the light green dots of Fig.~\ref{tDBLG(AB-AB)_Ez-Lso_band_gap_BZ}(e)] of mBZ. 
We observe similar band gap closing as we vary $\Delta$ in Fig.~\ref{tDBLG(AB-AB)_band_gap_BZ}(c) corresponding to a change in Chern number by $3$ in the $C_{\uparrow}^{K}$ phase diagram 
of AB-AB tDBLG [see Fig.~\ref{Phase_diag1}(c)]. 
Therefore, we conclude this section by noting that while most of the topological phase transitions we come accross in this work are supported by the Lifshitz like transitions (which is a well known phenomena in bilayer graphene~\cite{PhysRevB.98.165406}), here we report some other topological phase transitions (mainly because of the spin-orbit interaction present in the system) those are mediated by forming a rod-like (holding two pockets at the ends) shape in the mBZ, are not well established.  

\section{Summary and Conclusions}\label{Sec:VII}

To summarize, in this article, we investigate the electronic band structure and topological band properties of tDBLGs in the presence of intrinsic SOC and transverse electric field. To understand the topological properties of the system, we first explore the direct band gap closings in mBZ for the twisted case. Apart from the usual Lifshitz-like transition, we also observe one new kind of gap closing in the mBZ when SOC is present. Finally, we analyse the Chern numbers considering different valley and spin sectors and exhibit various topological phase diagrams. We emphasize how the increasing SOC strength is affecting the topological phase diagrams and introduces new topological phases with different Chern number when the electric field $\Delta$=0. We also find that while in the absence of intrinsic-SOC, the system is a QVHI, it turns into a QSHI and also remains a QVHI in the presence of the SOC for both AB-AB and AB-BA tDBLG. However, the parameter regime are different for the AB-AB and AB-BA case where they remain both QVHI and QSHI. Throughout the paper, we also compare the uDBLG results with that of the twisted one, to realize 
the effect of twisting on the topological band preperties of the system. 

Although in a previous article by Mohan et al. (2021)~\cite{Mohan2021} similar studies has been reported without SOC, however our model and findings are quite different. In our model, we introduce the spin degrees of freedom for the tDBLG system and demonstrate how the electronic band dispersion changes in the presence of intrinsic SOC. In contrast, the previous study incorporates a spinless model to describe the twisted system. Throughout the manuscript, we present a comparative analysis between untwisted and twisted double bilayer graphene, making it evident that the introduction of twist significantly modifies the phase diagram. The latter analysis is not provided in any previous work even without SOC. While the previous work reports only the valley-Hall insulator phase, our finding has identified a transition to the quantum spin-Hall insulator phase which is not possible in absence of spin degrees of freedom. Additionally, we establish the new topological phases with different Chern numbers originating due to intrinsic SOC. Importantly, new topological phases appear in the phase diagram which is completely trivial in absence of SOC. Hence, presence of SOC can give rise to topological phase starting from a non-topological one in tDBLG system.

While in our analysis, we deal with the tDBLG system, it is well known that graphene doesn't exhibit strong intrinsic spin-orbit interaction. However, our system can mimic a twisted double bilayer of silicene or similar type of materials like germanene or stanene, with intrinsic SOC present in them~\cite{twisted_silicene-Li2016,twisted_silicene-Han2023,PhysRevLett.130.196401}. We would like to note that there have been experimental realizations of twisted bilayers of germanene~\cite{Bampoulis_2024}, and we hope that our work will find relevance in this context in the near future if twisted double bilayer of germanene can be experimentally fabricated. However, the presence of buckling may affect the amplitude of the interlayer coupling $u$, $u^{{\prime}}$ and introduce a little quantitative variation in the phase diagram.

Note that, in our tDBLG model which encompasses a large parameter space, the lowest energy bands near charge neutrality may not always be isolated from the other valence and conduction bands while varying parameters such as the twist angle (\(\theta\)), gate voltage (\(\Delta\)), or SOC strength (\(\lambda_{\text{so}}\)). Consequently, calculating the total Chern number may lead to inaccuracies if the initial and final bands are not isolated from the other bands (\ie if a band gap is not maintained over the parameter regime). For this reason, as a precautionary measure, we prefer to compute the Chern number for all filled bands below charge neutrality rather than for a limited set of bands near charge neutrality. Moreover, as we increase the momentum cutoff, that adds more basis vectors to the model Hamiltonian, thereby increasing the number of bands. Using this higher momentum cutoff, we recalculate the Chern number and establish that it remains qualitatively unchanged despite the significant increase in the number of bands. Thus, one can infer that  the bands far from the Fermi energy do not contribute to the Chern number.


In conclusion, the presence of SOC turns tDBLG into a QSHI and provides an extra controlling knob apart from the gate voltage and twist angle for realizing new topological phases (both AB-AB and AB-BA case). Also, it has been emphasized that how the twisted case differs from the results of the uDBLG. Finally, we mention that a different kind of gap-closing transition is observed in presence of SOC, corresponding to the topological phase transitions.

\subsection*{Acknowledgments}
K.B. acknowledges Arnob Kumar Ghosh for stimulating discussion. K.B. and A.S. also acknowledge SAMKHYA: High-Performance Computing Facility provided by Institute of Physics, Bhubaneswar, for numerical computations.


\appendix
\section{Effect of sublattice dependent intrinsic SOC}
\label{AppA}
\vskip +0.2cm
\begin{figure*}[t]
	\centering
	\includegraphics[width=0.4\textwidth]{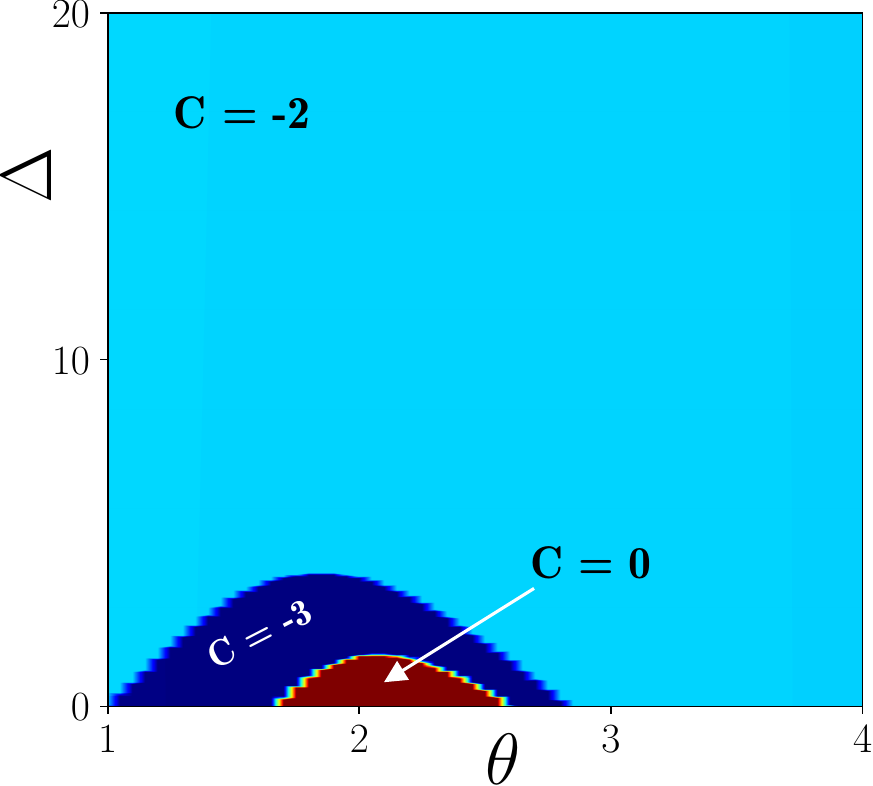}
	\caption{Topological phase diagram for the Chern number corresponding to the AB-AB tDBLG near valley-$K$, is depicted in the plane of gate-voltage ($\Delta$) and twist angle ($\theta$).  
	We choose the sublattice dependent intrinsic SOC, $\lambda_{{\rm{I2}}}$ = 10 meV and $\lambda_{\rm{I1}} = 9$ meV for sublattice ${\rm{A}}$ and sublattice ${\rm{B}}$, respectively.
	}
	\label{Lso11_Lso10_phase_diag}
\end{figure*}
In this appendix, we explore the effects of sub-lattice-dependent intrinsic SOC. To investigate this, we incorporate two sub-lattice-dependent intrinsic SOC parameters into our model Hamiltonian: $\lambda_{{\rm{I2}}}$ for sublattice ${\rm{A}}$ and $\lambda_{{\rm{I1}}}$ for sublattice ${\rm{B}}$, as mentioned in~\cite{Sub-lattice_dep_soc1, Sub-lattice_dep_soc2}. 
In Fig.~\ref{Lso11_Lso10_phase_diag}, we present a phase diagram of the Chern number near the valley $K$ correnponding to AB-AB tDBLG. 
Comparing this with the phase diagram for $\lambda_{so} = 10$ meV near valley $K$ and the spin-down sector of the AB-AB tDBLG [Fig.~\ref{Phase_diag1}(e)], we observe qualitatively similar results. Although the inclusion of sub-lattice-dependent intrinsic SOC makes the model more realistic, however it does not significantly affect the topological properties of the system.


\bibliography{bibfile}{}


\end{document}